\documentclass[twocolumn,times,tighten]{aastex63}
\usepackage{color}
\usepackage{enumitem}
\usepackage{hyperref}
\usepackage{verbatim}
\usepackage{amsmath,amstext,mathrsfs}
\usepackage[all]{hypcap} 
\usepackage{afterpage}    
\usepackage{marginnote}
\usepackage{makecell}
\usepackage{subfigure}
\usepackage{xfrac}

\DeclareFontFamily{OMS}{oasy}{\skewchar\font48 }
\DeclareFontShape{OMS}{oasy}{m}{n}{%
         <-5.5> oasy5     <5.5-6.5> oasy6
      <6.5-7.5> oasy7     <7.5-8.5> oasy8
      <8.5-9.5> oasy9     <9.5->  oasy10
      }{}
\DeclareFontShape{OMS}{oasy}{b}{n}{%
       <-6> oabsy5
      <6-8> oabsy7
      <8->  oabsy10
      }{}
\DeclareSymbolFont{oasy}{OMS}{oasy}{m}{n}
\SetSymbolFont{oasy}{bold}{OMS}{oasy}{b}{n}

\DeclareMathSymbol{\smallleftarrow}     {\mathrel}{oasy}{"20}
\DeclareMathSymbol{\smallrightarrow}    {\mathrel}{oasy}{"21}
\DeclareMathSymbol{\smallleftrightarrow}{\mathrel}{oasy}{"24}

\newcommand*{\re}{}

\shorttitle{Measuring B-Field Strength}
\shortauthors{Lazarian, Yuen \& Pogosyan}

\submitjournal{ApJ}

\begin{document}

\title{Obtaining magnetic field strength using differential measure approach and velocity channel maps}

\author[0000-0002-7336-6674]{Alex Lazarian}
\affiliation{Department of Astronomy, University of Wisconsin-Madison, USA}
\affiliation{Korea Astronomy and Space Science Institute, Daejeon 34055, Republic of Korea}
\email{alazarian@facstaff.wisc.edu}

\author[0000-0003-1683-9153]{Ka Ho Yuen}
\affiliation{Department of Astronomy, University of Wisconsin-Madison, USA}
\email{kyuen@astro.wisc.edu}

\author[0000-0002-7998-6823]{Dmitri Pogosyan}
\affiliation{Department of Physics, University of Alberta, Edmonton, Canada}
\affiliation{Korea Institute for Advanced Studies, Seoul, Republic of Korea}
\email{pogosyan@ualberta.ca}

\begin{abstract}
We introduce two new ways of obtaining the strength of plane-of-sky (POS) magnetic field by simultaneous use of spectroscopic Doppler-shifted lines and the information on magnetic field direction. The latter can be obtained either through polarization measurements or using the velocity gradient technique. We show the advantages that our techniques have compared to the traditional Davis-Chandrasekhar-Fermi (DCF) technique of estimating magnetic field strength from observations. The first technique that we describe in detail employs structure functions of velocity centroids and structure functions of Stokes parameters.  We provide analytical expressions for obtaining magnetic field strength from observational data. We successfully test our results using synthetic observations obtained with results of MHD turbulence simulations. We measure velocity and magnetic field fluctuations at small scales using two, three and four point structure functions and compare the performance of these tools. We show that, unlike the DCF, our technique is capable of providing the detailed distribution of POS magnetic field and it can measure magnetic field strength in the presence of both velocity and magnetic field distortions arising from external shear and self-gravity. The second technique applies the velocity gradient technique to velocity channel maps in order to obtain the Alfven Mach number and uses the amplitudes of the gradients to obtain the sonic Mach number. The ratio of these two Mach numbers provides the intensity of magnetic field in the region contributing to the emission in the channel map. We test the technique and discuss obtaining the 3D distribution of POS galactic Magnetic field with it. We discuss the application of the second technique to synchrotron data. 
\end{abstract}


\section{Introduction}

The role of magnetic fields in astrophysics is difficult to overestimate. Magnetic force is the second most important force in the present day Universe after gravity. The magnetic field plays an important role at different stages of star formation (e.g. \citealt{1956MNRAS.116..503M,2006ApJ...647..374G,2006ApJ...646.1043M,2007IAUS..243...31J}). In view of astrophysical flows with large Reynolds numbers the magnetic fields are turbulent (see  \citealt{2004ARA&A..42..211E,MO07,2016ApJ...824..113X,2016ApJ...832..199X}). The evidence of turbulent magnetic field is coming from observations of density structure of the interstellar medium (e.g. \citealt{1995ApJ...443..209A,CL09}) and velocity fluctuation studies \citep{1981MNRAS.194..809L,2004ApJ...615L..45H,2010ApJ...710..853C}.

\cite{1951PhRv...81..890D} and \cite{CF53} proposed the technique (Davis-Chandrasekhar-Fermi technique, henceforth DCF technique) that allows to estimate the magnitude of the mean magnetic field in interstellar medium and molecular clouds by measuring both the variations of the observed magnetic field directions and the dispersion of velocities. The technique has been revised and improved by the community (e.g. \citealt{2001ApJ...561..800H,2004Ap&SS.292..225C,2004ApJ...616L.111H,2006Sci...313..812G,Fal08}), but the foundations of the technique stayed the same. In particular, the DCF assumes that the perturbations of magnetic field direction arise from the collection of Alfven waves at all scales with waves at the largest scale dominating the observed variations. This, however, is not true for most of the astrophysical settings with large scale magnetic field fluctuations being affected by the factors not related to turbulence, e.g. by gravity. As a result, the accuracy of the DCF technique is low.

Improvements of the DCF technique was taken care first by replacing the dispersion of polarization angles to the structure functions of it in \cite{2009ApJ...696..567H}. The authors used the structure functions of magnetic field angle introduced in \cite{Fal08} and discussed how the dispersion of magnetic field can be obtained from observations.  \cite{2009ApJ...696..567H} models the structure function of magnetic field angles (not polarization angles) by the first two terms of its Taylor expansion $SF\{\phi\}({\bf R}) \sim b^2 + m^2 R^2$ to obtain the ratio between the turbulent-to-regular magnetic field strength. The calculations, however, assumed that the correlation length scale of turbulence is smaller than the separation between the line of sights at which the structure function is calculated. This assumption of very small scale turbulence was applied further in the subsequent study \citep{2009ApJ...706.1504H}. This assumption, as we discuss further, is not applicable for interstellar studies as there is good evidence that we resolve turbulence both in diffuse media (see \citealt{C10}) and in molecular clouds \citep{2009ApJ...707L.153P,2011ApJ...733..109H}. Therefore it is advantageous to explore what we can get using the small scale differential measures that are influenced mostly by turbulent motions. 

In a separate development, improvements on the dispersion of velocity has not been considered until the work by \cite{CY16} and the subsequent works \citep{Cho17,YC19,Cho19}. \cite{CY16} discuss the origin of magnetic field strength overestimation when using the DCF technique due to the multiple sampling of largest turbulent eddies along the line of sight. A suggestion of replacing the velocity dispersive measure from the velocity line width $\delta v_{los}$ to the dispersion of velocity centroids $\delta C$ was tested in \cite{CY16} and show that the replacement of velocity centroid correctly estimate the mean magnetic field strength on the plane of sky. The problem of dealing with the large scale variations of magnetic field while statistically determining the magnetic field strength was addressed in \cite{Cho19} by replacing the dispersion of velocity centroid to its multi-point structure function variants. 

A natural improvement of the DCF technique would be replacing both $\delta \phi$ and $\delta v$ by their structure functions. This came from the energy balance of the Alfven waves, namely the Alfven relation. The reason of why structure function of the observables are crucial in estimating magnetic field strength can be understood as follows assuming we are having an idealized fully driven incompressible magnetized turbulence with a constant density $\langle \rho \rangle$ :We can conjecture that the three-dimensional structure functions for both magnetic field and velocities be:
\begin{equation}
    SF_{2,3D}\{B\}({\bf r}) = 4\pi \langle \rho \rangle SF_{2,3D}\{v\}({\bf r})
    \label{eq:sf_conjecture}
\end{equation}
where $SF_{2}\{*\}$ denotes the 2nd order structure function for the variable $*$ and ${\bf r} = (x,y,z)$ the three-dimensional vector. In incompressible turbulence this conjecture makes sense since we expect magnetic fields and velocities are driven coherently. The structure function formalism would be very useful in the aspect of theoretical point of view since \cite{LP12} and \cite{KLP17a} tackle what physical properties are stored in the polarization angle and velocity structure functions respectively (See also \citealt{LP08,LP16}).

In what follows we propose a new technique that is based on the both the structure function measurements of both velocities and polarization angles from dust grain alignments (see \citealt{Aetal15} for a review) at small scales.  In \S \ref{sec:strength} we review the foundation of the CF technique and its limitations. In \S \ref{sec:input} we review the required input for the DCF technique. In \S \ref{sec:theory} we shall discuss the theoretical formulation of the differential measure method that allows the estimation of magnetic field strength in small scales. In particular, we developed a detailed formulation based on the respective structure function analysis \citep{LP08,LP12,LP16,KLP17a} in estimating the magnetic field strength when we have different ratios of MHD modes. In \S \ref{sec:method} we discuss our numerical methods in testing \S \ref{sec:theory}. In \S \ref{sec:num} we perform our numerical tests and provide the recipe in applying \S \ref{sec:method} in observations. In \S \ref{sec:multipoint} we discuss the potential use of the multipoint statistics in applying our new differential measure method. In \S \ref{sec:comparison} we compare our technique to other viable magnetic field strength estimation techniques.  We discuss the possible application of the Velocity Gradient observables in acquiring magnetic field strength in \S \ref{sec:VGTDMA}. In \S \ref{sec:ach} we discuss the achievements and the existing limitations. We discuss the potential applications of our technique in \S \ref{sec:discussion} and \S \ref{sec:conclusion} we summarize our work.

\section{Davis-Chandrasekhar-Fermi Technique and its Limitations}
\label{sec:strength}

\subsection{Basics of the DCF approach}

In Alfvenic motions the magnetic field fluctuations and those of the velocity are directly related through the averaged Alfven relation \citep{1942Natur.150..405A}:
\begin{equation}
\delta B=\delta v (4\pi \langle \rho\rangle)^{1/2},
\label{eq:alf}
\end{equation}
where $\langle \rho\rangle$ is the mean density, which can be obtained through independent observations. Therefore, by measuring $\delta v$ one can obtain the strength of magnetic field perturbation. These perturbations induce the deviations of the underlying field $B_{mean}$ by an angle $\theta$, which is $\delta \theta \approx \delta B/B_{mean}$. As a result, if one can measure $\theta$ it is possible to evaluate the strength of the underlying magnetic field in the system $B_{mean}$.

This simple physical mechanism is behind the DCF technique \citep{1951PhRv...81..890D,CF53} that uses global dispersion of magnetic field directions $\delta \theta$ and the velocity dispersion $\delta v$. The deficiency of the DCF technique is that it is sensitive to large scale magnetic field distortions that do not arise from turbulent velocities as well as to large scale shear. 

The Alfvenic Mach number is connected to the DCF technique since $M_A \sim \delta \theta$. As we shall discuss in \S \ref{sec:input} \& \S \ref{sec:discussion} that there are several techniques to find $M_A$. If the Alfven Mach number is known and less than unity , it is possible to obtain the mean magnetic field strength from the relation $M_A=\frac{\delta B}{B_{mean}}$. Assuming that the relation of $\delta B$ and the measured linewidth $\delta B$ obey the Alfvenic relation, i.e. see Eq.(\ref{eq:alf}) one can express the mean magnetic field strength as 
\begin{equation}
B_{mean}= C_1 \sqrt{4\pi \langle \rho \rangle} \frac{\delta v}{M_A}
\label{eq:BB}
\end{equation}
where $\delta V $ is the 3D velocity dispersion, which is to be determined from observations, while $C_1$ is an adjustable factor, reflecting the ambiguities associated with this simplified approach. 

We took into account that it is the plane of sky (POS) mean component of magnetic field that is being explored with the technique. In the time being we shall assume the total mean field are completely resides into the POS, then we can simply replace $B_{mean}$ to $B_{POS}$. As only line of sight (LOS) velocity $\delta v_{los}$ is available in observation, for the practical use of the Eq. (\ref{eq:BB}) the velocity dispersion $\delta v$ there should be associated with $\delta v_{los}$, i.e.  $\delta v=C_2\delta v_{los}$ , where $C_2$ is a coefficient that relates the dispersions of the turbulence POS velocities with the available LOS ones. For an uniformly distributed Alfven wave that moves along the mean magnetic field line, $C_2=\sqrt{2}$ due to the 2 degrees of freedom the Alfven wave enjoy. The coefficient $C_2$ grows up when the angle between the mean magnetic field direction and the line of sight is smaller. In the limiting case of magnetic field parallel to the line of sight, $C_2$ is not defined,  as no line of sight velocities can be associated with the Alfven motions.

\subsection{DCF and MHD turbulence}

It is well known that the actual interstellar turbulence is different from the superposition of Alfvenic waves that is discussed in the pioneering studies. 
The simplest approximation is the incompressible turbulence. The extensive studies of this regime of turbulence during last two decades revealed a few  distinct regimes (see \citealt{2019tuma.book.....B}) which we describe below. We have a short summary of properties of MHD turbulence in Appendix A.

\subsubsection{Super-Alfvenic Turbulence}

If the velocity at the injection scale $L_{inj}$ is larger than Alfven velocity, the turbulence is super-Alfvenic and $M_A>1$ The hydrodynamic motions easily bend magnetic field at the injection scale and the observed distribution of magnetic field is random. However, at as turbulence cascades the motions at the scale $l_A=L_{inj}M_A^{-3}$ become Alfvenic and the are dominated by magnetic forces. Starting from this the amplitudes the velocity perturbations and the amplitudes magnetic field perturbations are related to the magnetic field strength.

It is clear that the DCF approach is not applicable to super-Alfvenic turbulence. Indeed, the dispersion of magnetic field directions on the large scale is determined by hydrodynamic motions that are marginally affected magnetic field strength. 

\subsubsection{Sub-Alfvenic Turbulence}

In the opposite case, i.e. when the injection velocity is less than the Alfven velocity, the turbulence is sub-Alfvenic and magnetic fields are strong enough to affect the magnetized fluid motions from the injection scale. Therefore again the amplitudes of velocities and magnetic field perturbations are related to the magnetic field strength.

It is also clear that even for sub-Alfvenic turbulence the DCF approach that ignores the actual properties of magnetized turbulence cannot provide accurate magnetic strength measurements. 

\subsubsection{Applying DCF in compressible turbulence}

One should remember, however, that even in the case of incompressible MHD turbulence Alfvenic turbulence is not acting alone. The fluctuations of magnetic field compression which are the degenerate limiting case of slow waves, pseudo-Alfven waves, exist in this case. The scaling properties of pseudo-Alfven and Alfven modes are similar, as the Alfven modes shear pseudo-Alfven perturbations and impose their scaling on them \citep{GS95}. 

The situation gets more complex in the compressible MHD turbulence. In compressible media MHD turbulence can be decomposed into Alfven, slow and fast modes. These modes cascade, but the properties of them are very different. For instance, Alfven modes are are mostly responsible for the deviations of magnetic field directions (see \citealt{LV99}).  
\subsection{Linewidth and the dispersion of velocity centroids}

Some issues with the DCF are so self-evident that it is a bit surprising that in its simplest incarnation the approach safely survived till now. The most obvious problem is related to the use of the linewidth in Eq. (\ref{eq:BB}). The measured linewidth does not change if the emitting region extends for more than one injection scale $L_{inj}$ along the line of sight. At the same time, as it was correctly pointed out by \cite{CY16}, the magnetic fields on the scale larger than $L_{inj}$ are being summed up in a random walk manner. The authors provided their solution that we briefly discuss in Appendix \ref{app_CF}.

\cite{CY16} uses the dispersion for the velocity centroids $\delta C$ (see Eq.\ref{centroid}) for the expression of velocity centroid) instead of the linewidth that also can be obtained from observations. The elementary transformation to the line of sight integration involve the Jacobian change according to $\rho_v dv=\rho (z) dz$, where $\rho (z)$ is the density of emitters in along the line of sight. Thus the integral in the numerator is equal to $\int_{\cal L} v(z) \rho (z) dz$, where ${\cal L}$ is the integration length along the line of sight/ In the limit of constant density, i.e. $\rho(z)=\rho$, provides the mean velocity along the line of sight multiplied by $\rho {\cal L}$. The integral in the denominator provides the column density $\rho {\cal L}$ which is proportional to intensity of measured radiation. Therefore, for incompressible turbulence $C$ provides the turbulent velocity averaged along the line of sight. The dispersion of $C$ at the injection scale are compared to the dispersion of polarization directions to obtain the strength of the mean magnetic field as suggested in \cite{CY16}. 

In the \cite{CY16} approach both the velocities and magnetic fields are summed up in the same way along the line of sight (see Appendix \S \ref{app_CF}) and therefore the modified technique is applicable to ${\cal L}>L_{inj}$ cases. Nevertheless, the modification of the technique shares with the DCF approach the limitations related to the nature of magnetic fluctuations.

\subsection{Other limitations of the DCF and attempt to improve the technique}

Apart from using the over-simplified model of magnetic and velocity fluctuations in the interstellar medium, the DCF approach has additional deficiencies. For instance, determining the velocity dispersion arising in realistic astrophysical settings can be problematic for the DCF. In many cases the line broadening is affected by the shear not related to turbulence. A typical example is diffuse HI, for which the non-thermal broadening mostly arise from galactic rotation and therefore the measured line widths are not meaningful within the DCF approach. 

Determining the magnetic field dispersion can also be problematic for the DCF. Self-gravity presents a serious problem for the DCF technique as it induces the dispersion of magnetic field directions that does not related to the effect of turbulence. Separating of the two withing the DCF approach that uses the global dispersions of the magnetic field directions may not be possible.   

There have been numerous attempts to improve the accuracy of the DCF technique, most of them based on numerical testing with attempts to adjust the factor $C$ in Eq. (\ref{eq:BB}) (see e.g. \citealt{CY16}) or extend the technique to larger range of $M_A$ \citep{Fal08}. We also mentioned approach in \citep{CY16} (see Appendix \ref{app_CF}). Nevertheless, all these studies have not attempted to change the nature of the DCF technique, namely, its use of global dispersions of the measures employed. 

In this paper we advocate the technique that uses the same observables, but in a different way. The new technique that employs the differential measures of increments of $\delta \theta$ and $\delta v$ at small scale where the contribution of global inhomogeneities is mitigated. To distinguish the two approaches we will use the term Differential Measure Approach (henceforth, DMA). Note, that as we explain later in \S \ref{sec:comparison} our approach is very different from that proposed in \cite{2009ApJ...696..567H} and developed in subsequent publications.

\section{The input information for obtaining magnetic field strength}
\label{sec:input}

As we discuss in \S \ref{sec:strength} the DCF technique \citep{1951PhRv...81..890D,CF53} is based on the assumption that the observed fluctuations are Alfven waves. In this simplified model the amplitude of magnetic fluctuations for a given velocity perturbation depends on the strength of the magnetic field. Thus, it was suggested that the amplitude of velocity can be measured due to the Doppler shift and the magnetic field perturbation can be measured with dust polarization and this would provide the magnetic field strength. 

The techniques that we discuss in the paper similarly use the information about the magnetic field and the non-thermal velocities, but in a different way.  Below we list the measures that provide the required information. 

\subsection{Velocity centroids and channel maps as an observable of velocity information}
The velocity information on astrophysical turbulent volume is available from observations in the form of Position-Position-Velocity (PPV) cubes where intensity $I(\mathbf{X},v)$ that is the measure of the plane of sky coordinate ${\bf X}$ and the Doppler-shifted velocity $v$. There can be different ways of study of this quantity. One can study the intensities in channel maps, integrating  $I(\mathbf{X},v)$ over the thickness of the velocity channel. This is the basis of the Velocity Channel Analysis (VCA) technique introduced in  \citep{LP00}. However, dealing with the DMA we shall focus on the velocity moments of $\rho(\mathbf{X},v)$, e.g. the normalised first moment, i.e. the velocity centroid, 
\begin{equation}
C(\mathbf{X}) \propto \int_a^b dv v \rho(\mathbf{X},v)/ \int_a^b dv \rho(\mathbf{X},v) ,
\label{centroid}
\end{equation}
where depending on the choice of the integration limits one can get different measures. For instance, integrating over the entire spectral line width one gets a measure known as a velocity centroid. If the integration limits are chosen over a part of the line, we are dealing with the {\it reduced centroids} \citep{LY18a}. For the incompressible fluid the velocity centroids provide the value of velocity averaged along the line of sight. The reduced centroids are valuable for probing turbulence in the presence of galactic rotational curve. Then, in the incompressible limit the reduced centroids provide the estimate of the mean velocity over a selected part of the galactic media.

In addition to velocity centroids, fluctuations of intensity in velocity channels carry the information about the turbulent velocity field. The use of velocity channels is based on theory of turbulent fluctuations in Position-Position-Velocity (PPV) space introduced in \cite{LP00} and elaborated in subsequent publications \citep{LP06,KLP16}.

\subsection{Tracing magnetic fields with polarization}

\noindent {\bf Stokes parameters:}
In polarization observations the magnetic field is measured using the Stokes parameters for synthetic observations are given by
\begin{equation}
\label{eq.stokes}
    \begin{aligned}
    Q&\propto\int dz n \cos(2\theta)\sin^2\gamma_{inc}\\
    U&\propto\int dz n \sin(2\theta)\sin^2\gamma_{inc} \\
    \theta_{pol} &= \frac{1}{2} \tan_2^{-1}(U/Q)
    \end{aligned}
\end{equation}
where $n$ is the number density, $\theta,\gamma_{inc}$ are the POS positional angle of the magnetic field and the inclination angle of magnetic field with respect to the line of sight, respectively. 
\linebreak

\noindent {\bf Polarization from aligned dust:}
Dust polarization arises from emission of non-spherical grains aligned with long axes perpendicular to the ambient magnetic field (see \citealt{Aetal15}). Similarly, polarization of starlight arises from the differential extinction by aligned grains. The processes of dust alignment is generally believed to happen due to radiative torques (RATs) (see \citealt{1976Ap&SS..43..257D,1996AAS...189.1602D} ). The theory of the RAT alignment have is based on the analytical model in \cite{2007MNRAS.378..910L} and further studies e.g. in \cite{2008MNRAS.388..117H,2016ApJ...831..159H}.

The RAT alignment theory at its present form (see \citealt{2019ApJ...883..122L}) can account for the major observational features of grain alignment. In particular, in typical conditions of diffuse ISM the silicate grains are nearly perfectly aligned, while in dense molecular clouds the degree of alignment depends on the grain illumination mostly by embedded stars. In other words, the existing grain alignment theory can evaluate in what conditions one should expect the polarization arising due to the aligned dust to trace magnetic fields.  With more polarization measurements obtained using starlight and with more distances to stars measured there is a possibility to trace magnetic field in 3D.
\linebreak

\noindent {\bf Goldreich-Kylafis Effect:} \citeauthor{1981ApJ...243L..75G} (\citeyear{1981ApJ...243L..75G,1982ApJ...253..606G}, henceforth GK) effect provides a viable way of tracing magnetic fields in molecular clouds. The polarization arises due to the differences of the radiation transfer in the media with anisotropies or shear. The resulting polarization is either parallel or perpendicular to the magnetic field. In spite of this ambiguity, the effect has been successfully employed to trace magnetic field structure of molecular clouds \citep{li}. Combining GK with velocity gradients one can remove the 90 degree ambiguity in the magnetic field direction. 
\linebreak

\noindent {\bf Ground State Alignment:} A promising development in terms of magnetic field tracing is presented by the atomic/ionic ground state alignment (GSA) effect suggested and quantified for use in astrophysical conditions by \citep{2006ApJ...653.1292Y,2007ApJ...657..618Y,2008ApJ...677.1401Y,2012JQSRT.113.1409Y}. The GSA employs atoms/ions with fine and hyperfine split levels. The atoms/ions get aligned in the ground or metastable state by external anisotropic radiation. The Larmor precession in the ambient magnetic field re-aligns the atoms/ions imprinting its direction on polarization. The atoms/ions stay in ground or metastable state long and thus they can trace very weak magnetic fields.  The effect has been recently confirmed with observations \citep{2019arXiv190308675Z}, opening a wide avenue of applying it for tracing magnetic fields in various environments. The difference in distribution of atoms and conditions for atomic alignment in space provides a way to get the 3D distribution of magnetic field in diffuse medium. The technique is especially interesting for probing magnetic field direction near bright sources. 
\linebreak

\subsection{Tracing magnetic field from velocity gradients}

In a recent series of papers we introduced velocity gradients as a way of tracing magnetic field (see \citealt{GCL17,YL17a,YL17b,Letal17,LY18a,survey}). The physical explanation why velocity gradients in diffuse media are perpendicular to the LOS projected magnetic field is routed in the theory of MHD turbulence \citep{GS95} and turbulent reconnection \citep{LV99}.

In particular, the theory of magnetic turbulent reconnection \citep{LV99} predicts that the turbulent motions perpendicular to the magnetic field are not constrained by the back-reaction of magnetic field. This presents the favorable way of turbulent cascading with most energy concentrated in the form of eddies perpendicular to the {\it local} direction of magnetic field.\footnote{A common misconception about the MHD theory is related to the fact that the concept of local direction of magnetic field is not a part of the original \cite{GS95} idea. As we discuss this concept naturally follows from turbulent reconnection, which is proved in the subsequent numerical studies \citep{CV00,MG01,CL02}.}  The notion of "local magnetic field of eddies" is the key for understanding how the gradient technique works. Indeed, if the rotation of turbulent eddies is aligned with magnetic field, then the gradients of velocity amplitudes are perpendicular to the magnetic field and therefore they can trace the magnetic field direction. As the magnetic field reconnects in one eddy turnover time  \citep{LV99} the eddy motions are Kolmogorov-like with the scaling of turbulent velocities $v_l\sim l^{1/3}_{\bot}$, where $l_\bot$ is eddy diameter perpendicular to local direction of magnetic field. As a result, the gradients of velocity amplitude scale as $v_l/l\bot\sim l^{-2/3}$, meaning that the maximal gradients are produced by the smallest resolved eddies. Due to this scaling, regular sharing motions do not affect the velocity gradient measurements.
\linebreak

\noindent {\bf Tracing B-fields with gradients:} A formal discussion of the velocity gradient technique is provided in \cite{LY18a}. There it is shown that the gradients arising from Alfven and slow modes are perpendicular to the magnetic field, while the gradients arising from fast modes are parallel to magnetic fields. Studies of compressible MHD turbulence show that the dominant contribution in most cases arises from Alfven and slow modes.
 
Velocity gradients present a possibility of 3D studies if different molecular lines are used. Indeed, different molecules are produce and survive at different depth in molecular clouds. This opens a possibility of studying magnetic fields in molecular clouds at different depths \citep{YL17b,velac} In addition, galactic rotation provides a way to probe magnetic field at different distances from the observer \citep{CL18}. Note, that due to the galactic rotation, velocity gradients can sample magnetic fields in many more clouds in the galactic disc compared to far infrared polarimetry. For the latter the confusion of emission from different clouds along the line is sight is detrimental. 
\linebreak

\noindent {\bf Dispersion of Velocity Gradient Orientation:} \cite{2018ApJ...865...46L} discuss the possibility of obtaining $M_A$ through the Velocity Gradient Technique \citep{GCL17,YL17a,YL17b,Letal17,LY18a}. The gradients of spectroscopic observables in diffuse interstellar media are test both numerically and observationally that they are perpendicular to the local magnetic field directions. Moreover, \cite{2018ApJ...865...46L} showed that the dispersion of the velocity gradient orientation is correlated to the local Alfvenic Mach number. Employing this technique \cite{survey} is possible to estimate the magnetization of a number of molecular clouds on the sky and obtain previously unachievable magnetic information on a high velocity cloud hid behind the galactic arm. 
\linebreak

\section{New Technique: Differential Measure Analysis}
\label{sec:theory}
\subsection{Simplified approach}
\label{subsec:simplified}

Consider the variations of the observed magnetic field direction within a volume with size ${\cal L}$ measured along the line of sight and the turbulence injection scale $L_{inj}$. The variations of the magnetic field angle can be characterized by 
\begin{equation}
\tan\delta\theta \approx \frac{\int \delta B dz}{\int B_{POS} dz}
\label{eq:tantheta}
\end{equation}
where the integration is done along the line of sight and 
where, without losing generality, we assumed that $\delta B$ is measured along the line of sight { and } perpendicular to the mean magnetic field $B_{POS}$ in the plane of the sky. Naturally, for sufficiently small $ \int \delta B dz/\int B_{POS} dz$  an approximation $\tan \delta \theta \approx \delta \theta$ is valid (See \citealt{Fal08}). However, in this study we do not need to use this approximation. 

If we are interested in the spatial variations of the observed magnetic field directions at the scale $l$, those can be obtained using the {\bf second-order structure functions ($SF$)} of the polarization angle $\phi$\footnote{We are here to use $\theta$ to denote the magnetic field angle and $\phi$ as the polarization angle because there is a subtle difference between them in some special geometry of magnetic field lines. See \cite{LY18a} for a discussion.}:
\begin{equation}
SF_{2D}\{\phi\}({\bf R} )= \langle [\phi ( {\bf X}+{\bf R}) -\phi( {\bf X})]^2\rangle_{{\bf X} }
\label{d_theta}
\end{equation}
where ${\bf X}$ is a two dimensional vector Plane of Sky (POS), $\langle...\rangle_{{\bf X }}$ denotes an ensemble averaging on the variable ${\bf X }$. For practical applications this means averaging for different ${\bf X}$ over the area $\gg l^2$. If within this area the $\int B_{POS} ds$ does not significantly change, the averaging in Eq. (\ref{d_theta}) amounts to averaging of the integrals of structure functions
\begin{equation}
SF_{2D}\{B\} ({\bf R })=\iint  SF_{3D}\{b\}({\bf r}) dz_1 dz_2
\label{st_turb}
\end{equation}
where $SF_{3D}\{b\}({\bf r})$ is the structure function of the POS magnetic field
\begin{equation}
SF_{3D}\{b\}({\bf r})=\langle [b_{turb} ({\bf x}+{\bf r})- b_{turb}({\bf x})]^2\rangle_{\bf x}.
\label{struc_b}
\end{equation}
with ${\bf x}$ be the 3D position vector; $z_1$ and $z_2$ denote the line of sights along which the integration of the structure function of 3D fluctuating magnetic field $b_{turb}$ is performed. 

In the system of the mean magnetic field, which is the only system that is available in the absence of 3D data, there is no scale-dependent anisotropy that is predicted in GS95 relations. The anisotropy at all scales is determined by the variations of the magnetic field direction at the injection scale, as it was demonstrated in \cite{CLV02}. Therefore, the same spectral slope of the fluctuations can be measured parallel and perpendicular to the mean magnetic field. In this situation, for the sake of simplicity, we will use structure functions averaged over the positional angle, which will make these functions only dependent on the line of sight distance $l$ separating the points. 

Observing that fluctuations of turbulent field are accumulated along the line of sight ${\cal L}$ in a random walk fashion one gets the structure function of the polarization angle $\phi$
\begin{equation}
SF_{2D}\{\phi\} (l) \approx SF_{3D}\{b\} (l) l {\cal L}
\label{eq:sf2d3d}
\end{equation}
where $l$ is the separation between the lines of sight. In statistical sense, the turbulence has the axial symmetry around the direction of mean magnetic field (see discussion in \citealt{LP12}). As a result, when we observe perpendicular to the magnetic field the eddies that have cross-section $l$ in the POS plane, have also the extension $l$ along the line of sight. These eddies are independent entities at the scale $l$ and therefore their summation happens in the random walk fashion. This provides the physical justification of Eq. (\ref{eq:sf2d3d}). 

In fact, the problem at hand has 3 scales - separation on the sky $l$, integration/cloud depth $\mathcal{L}$ and the injection scale $L_{inj}$, which is also the line-of-sight correlation length. So the answer is expressible via those three. If $\mathcal{L} \gg l$, we should get $SF_{2D}(l) \propto SF_{3D}(l) l \mathcal{L}$, so extra length factor is variable with $l$, which makes the slope steeper by unity as long as $l$ is sufficiently small compared to $L_{inj}$  and $\mathcal{L}$. This is the principal case that we are interested to explore in this paper.

Note, that in the limiting case of  $l\gg \mathcal{L}$ i.e we just take a narrow slice, we get $SF_{2D}(l) = SF_{3D}(l) \mathcal{L}^2$. This is a special case of studies when only a narrow surface area of the turbulent volume being proved by observations. This case can be realized in the presence of strong dust absorption as discussed in \cite{KLP18}. We do not discuss this case in the present work. 

On the contrary, the integration of the mean field proceeds in a regular way and therefore the cumulative effect of summing up its contributions is $B_{POS} {\cal L}$. As a result, the measured structure function is 
\begin{equation}
\sqrt{SF_{2D}\{\theta\}(l) }\approx \frac{SF_{3D}^{1/2}\{b\}}{B_{POS}}\sqrt{\frac{l}{\cal L}}.
\label{struc_main}
\end{equation}

For Alfvenic turbulence the fluctuations of velocity and magnetic field are symmetric. Therefore $v_{turb}=b_{turb}/\sqrt{4\pi \rho}$ , where $\rho$ is the plasma density. In terms of structure functions this means that the structure function of velocity:
\begin{equation}
SF_{3D}\{v\}(l)=\langle [v_{turb} ({\bf x}+{\bf l})- v_{turb}({\bf x})]^2\rangle_{\bf x}.
\end{equation}
is related to the structure function of magnetic field in Eq. (\ref{struc_b}) as 
\begin{equation}
SF_{3D}\{b\}=4\pi \langle \rho \rangle SF_{3D}\{v\}.
\label{struc_rel}
\end{equation}
where the averaging variable ${\bf x}$ is suppressed.

With observational spectral line data, one can measure the structure function of velocity centroids:
\begin{equation}
SF_{2D}\{C\} ({\bf R}) =\langle [C ({\bf X}+{\bf R})-C({\bf X})]^2\rangle_{\bf X},
\end{equation}
which presents the proxy of the structure function of the velocities, averaged along the line of sight. Due to this summing up of velocities procedure, the addition of velocity fluctuations happens similar  similar to summing up of magnetic perturbations $\delta b_{turb}$ that we deal with earlier. As a result, the summation process of the velocity fluctuations is a random walk process, i.e.
\begin{equation}
SF_{2D}\{C\}({l})\approx \int_{\cal L} SF_{3D}\{v\}\approx SF_{3D}\{v\}\frac{l}{\cal L}
\label{struc_centr}
\end{equation}
Combining Eqs. (\ref{struc_centr}), (\ref{struc_rel}) and (\ref{struc_main}), one gets the expression for the mean magnetic field:
\begin{equation}
B_{\bot}\approx f \sqrt{4\pi \langle \rho \rangle}\frac{SF_{2D}^{1/2}\{C\}(l)}{SF_{2D}^{1/2}\{\phi\}(l)}
\label{mean_B}
\end{equation}
where both $SF_{2;centroid}(l)$ and $SF_{2;\theta}(l)$ are available from observations and $f$ is constant of order unity that depends on the percentage of fundamental modes (see \citealt{CL02}) that compose the MHD turbulence. One can argue that for Alfvenic motions $f$ should be $\approx 1$ as this case the fluctuations of velocity and magnetic field are identical in amplitude. In our simplified approach $f$ is a factor that should be determined from numerical simulations. In general, can also have the dependence on the angle between the line of sight and the mean magnetic field direction.

We would like to stress that Eq. (\ref{mean_B}) is applicable to situations that the turbulence injection scale $L_{inj}$ is larger or smaller ${\cal L}$, as long as ${\cal L}\gg l$, the correlation length. The only requirement is that ${\cal L}$ should be the same for the calculations of $SF_{2D}^{1/2}\{C\}(l)$ and $SF_{2D}^{1/2}\{\phi\}(l)$. This requirement is automatically fulfilled if we use velocity gradients or ground state alignment are used to find $SF_{2D}^{1/2}\{\phi\}(l)$. The case of dust polarization requires more care to be sure that the polarization is collected from the same column of gas that contributes to the line emission. For instance, if the used line is 13CO, it is necessary to make sure that the column density of gas associated with CO emission is much larger than the column density of the of HI along the same line of sight. 

If turbulence is uniform and homogeneous Eq. (\ref{mean_B}) is equivalent to Eq. (\ref{d_theta}) as for $l\rightarrow\infty$ the structure functions get proportional to the total dispersion. However, in astrophysical situations, we have to deal with inhomogeneous samples for which differential measurements that reveal small scale inhomonogeneities are advantageous. We shall call the method of differential measures the {\bf Differential Measure Analysis (DMA)}.
 
The advantages of using the new DMA compared to DCF can be briefly summarized as follows:
\begin{itemize}
\item While dispersion of velocities and magnetic field directions that are employed by DCF are distorted by the linear large-scale shear, the structure functions used in the DMA  are not sensitive to it.
\item On large scales the structure of observed magnetic and velocity field is determined by gravity, outflows and other galactic processes determining the contours of individual molecular cloud, this is not a problem for the DMA that focuses only on the small scale differences in magnetic and velocity properties. 
\item Self-gravity induces additional distortions of magnetic field making the classical DCF approach not applicable. The DMA is expected to work in the case of the distorted magnetic field.
\end{itemize}
 
A clear illustration of the first point in the list above is that Eq. (\ref{mean_B}) is applicable to studies of magnetic field using the 21 cm line of atomic hydrogen. This line is broadened by both thermal motions and also galactic rotation, but one can still use structure functions of velocities using Reduced Velocity Centroids (RVCs) introduced in \cite{LY18a}. Using the gradients of the RVCs one can trace the distribution of magnetic fields as a function of distance from the observer (see \citealt{CL18}). As a result, Eq. (\ref{mean_B}) allows one to get the 3D distribution of $B_{POS}$ in the Galactic disk.

We note that if we apply Eq. (\ref{mean_B}) to the turbulence at large scale for $l$ comparable with the turbulence injection scale $L_{inj}$, we are getting not the DCF classical expression, but its generalization obtained in \cite{CY16}. We show in Appendix \ref{app_CF} that in this limit we can obtain can obtain magnetic for cases that ${\cal L}> L_{inj}$, which is beyond the domain of the traditional DCF formula.
 
\subsection{Detailed calculations}
\label{subsec:detailed}

DCF approach does not take into account the properties of MHD turbulence. Based on the idea of linear Alfven waves, it was assumed to be applicable to more realistic turbulent settings. Our estimates in the previous section went one step further by taking into account that turbulent eddies produce random walk when their contributions are summed up along the line of sight. However, the actual MHD turbulence is more than that. In \citeauthor{LP12} (\citeyear{LP12}, henceforth LP12) we described the statistics of magnetic fluctuations arising from MHD turbulence. In the subsequent study by \citeauthor{KLP17a} (\citeyear{KLP17a}, henceforth KLP17) the fluctuations of velocities have been described following the approach in LP12. These papers provide the basis for our detailed calculations. 
 
In what follows, we use the results of LP12 and KLP17 to have a derivation valid in the case of correlated magnetic field and velocity fluctuations, without reliance on the simplified considerations of random walk integrating over the line of sight that we used in the previous section. To do this we use "synchrotron polarization" formalism from Appendix of LP12 to describe the direction of magnetic field. Note, that the approach LP12 does not depend on the way we trace magnetic field. It can be synchrotron polarization or dust polarization, or velocity gradients etc since the mathematical structure of structure functions computed by the magnetic field directions traced by these methods exhibit the same behavior.

The strategy below can be literally summarized as follows: (1) We would first discuss what are a legitimate structure functions for velocity and magnetic field angles that could be measured observationally. (2) We then derive the expression of the structure function in 2D in relation to its 3D variant (3) We perform multipole expansion for each of the 2D structure functions according to \cite{LP12} for magnetic field angles and \cite{KLP17a} for velocity centroids. (4) We shall see how the ratio of the multipole terms of the structure functions would resemble the magnetic field strength of a given localized volume.

\subsubsection{Multipole expansion of Stokes Parameter structure functions}

For angle of polarization signal 
that traces the magnetic field direction (synchrotron, dust polarization, synthetic polarization from gradient maps)  we can quite generally write 
\begin{eqnarray}
\cos(2 \phi) &=& \frac{ \int dz (H_x^2 - H_y^2)}{\int dz (H_x^2 + H_y^2)}
\propto Q/I\\
\sin(2 \phi) &=& \frac{ \int dz 2 H_x H_y}{\int dz (H_x^2 + H_y^2)} \propto U/I 
\label{eq:stokes1}
\end{eqnarray}
where $I,Q,U$ are the Stokes parameters (See \S \ref{sec:num}) from which we can construct the structure function.\footnote{We drop here the degree of polarization $p$, since we are just interested in the structure functions of polarization angles.}
\begin{equation}
\begin{aligned}
&\left\langle \left( \frac{Q_1}{I_1}-\frac{Q_2}{I_2}\right)^2\right\rangle
+
\left\langle \left( \frac{U_1}{I_1}-\frac{U_2}{I_2}\right)^2\right\rangle\\
&= 2  \left\langle 1 - \cos(2(\phi_1-\phi_2)
\right\rangle \\
\end{aligned}
\label{eq:correct_dtheta}
\end{equation}
In the limit of small angle differences this structure function is proportional to one given by Eq. (\ref{d_theta}). In fact, in the case of small angle differences Eq.\ref{eq:correct_dtheta} reduces to $4\langle (\phi_1-\phi_2)^2\rangle$. However, it is a more general expression that is better defined from observations, and also applicable to the case
when angle fluctuations are large, e.g., when the Alfvenic Mach number is large.

If we assume that the denominator is dominated by the mean field, we get
\begin{equation}
2 \left\langle 1 - \cos(2 (\phi_1-\phi_2) \right\rangle
\approx \frac{ D_{QQ} + D_{UU}}{(\bar I)^2}
\label{eq:expand}
\end{equation} 
where, using the notation listed in Table \ref{tab:params} (See also \citealt{LP12}):
\begin{eqnarray}
\bar I = \overline{ \int dz (H_x^2 + H_y^2)} = 
 \mathcal{L} \sigma_{H_\perp}^2 
\label{eq:bari}
\end{eqnarray}
Following Appendix C \& D in \cite{LP12},
\begin{equation}
D_{QQ}+D_{UU} \approx 4 \mathcal{L} \sigma_{H_\perp}^2 \int dz ( D^+(R,z)-D^+(0,z))
\label{eq:Dqquu}
\end{equation}
thus
\begin{align}
D^\phi & \equiv \frac{1}{2}\left\langle 1 - \cos(2 (\theta_1-\theta_2) \right\rangle
\nonumber \\
& \approx \frac{ \int dz ( D^+(\mathbf{R},z) - D^+(0,z))}{\mathcal{L}  \sigma_{H_\perp}^2}
\label{eq:dphi}
\end{align}
We note, that this derivation assumed that the mean magnetic field dominates the perturbations, so in the same spirit one can replace the second moment by the square of the magnetic field, $\sigma_{H_\perp}^2 \approx \bar{H}_\perp^2$.

The statistics of centroids was recently discussed in \cite{KLP17a}. There, for the sake of theoretical convenience the definition of centroids was modified compared with the standard one given by Eq. (1). In particular, the numerator was divided not by the intensity at the given point, but by the mean intensity. A numerical study in \cite{EL05} shows that this change does not significantly alter the statistics of the centroids. At the same time, this significantly simplifies the analytical treatment of the centroids.

In multipole representation $\int dz D^+(\mathbf{R},z)$
has coefficients
\begin{align}
D^+_n = A_B^{(A,F,S)} C_n(m) R^{1+m}
\sum_{s=-\infty}^\infty \widehat{E}_s G_{n-s}^{(A,F,S)}(\gamma)
\label{eq:Dn+}
\end{align}
where the amplitude $A_B$ that has dimensions of 
$\left[H^2 L^{-m}\right]$ \footnote{We denoted with hat all non-obviously dimensionless quantities} appears in the definition of the power spectrum of a given turbulent mode
\begin{equation}
E(k,\mu = \hat{k}\cdot \hat{\bf B}) = A_B^{(A,F,S)} k^{-3-m} \widehat{E}(\mu)
\label{eq:espec}
\end{equation}
The amplitude can be related to the variance of the magnetic field perturbation as
\begin{equation}
A_B^{A,F,S} L_{inj}^m = \left\langle \delta B^2 \right\rangle \times \frac{(2\pi)^2}{\int_1^\infty d \ln{k} k^{-m} \int_{-1}^1 d\mu  \widehat{E}(\mu)} 
\end{equation} 
where $L_{inj}$ is the energy injection scale. Accurate model involve smooth truncation of the power spectrum at this scale, rather than a sharp cutoff at dimensionless wavenumber satisfying $k L_{inj}=1$, but the dimensionless integral factors will anyway drop out from the subsequent consideration. 

For the structure function of angle fluctuations we therefore obtain
\begin{align}
D^\phi_n = \frac{A_B^{A,F,S}}{\mathcal{L} \bar H_\perp^2}
C_n(m) R^{1+m}
\sum_{s=-\infty}^\infty \widehat{E}_s G_{n-s}^{(A,F,S)}(\gamma)
\label{eq:phimultipole}
\end{align}
where effective $L_{inj}/\mathcal{L}$ factor expresses the suppression of structure function amplitude due to random walk in the inertial range that starts with
$L_{inj} < R < \mathcal{L}$ range. Note that the \cite{LP04} and \cite{KLP16} formalism uses $k_z=0$ approximation for evaluating $z$ integral on z-coordinate differences along two line of sights,
assumes that the integration range exceeds $k_z^{-1}$ of any scale of interest. Thus it is not applicable
near $L_{inj}$ if $L_{inj} > \mathcal{L}$ since $ \mathcal{L}^{-1} > k_z \sim L_{inj}^{-1}$ will not average out, so we assume that the
integration depth exceeds the injection scale.

\subsubsection{Multipole expansion of velocity centroid structure functions}
The subsequent step is to evaluate the structure function in the nominator via the structure function of velocity 
centroids which has a very similar behaviour, given that
the magnetic field and velocity scales in the same way.
Indeed the multiple moments of the structure function of centroids \citep{KLP17a}, normalized by the mean column intensity of the gas along the line of sight $\bar I = \epsilon \mathcal{L} \bar\rho$,
have the form
\begin{align}
\widetilde{\mathcal{D}}_n &=\mathcal{D}_n/ (\epsilon \bar{\rho} \mathcal{L})^2 = \nonumber \\ 
&=(L_{inj}/\mathcal{L}) A_v C_n(m)\sum_{p=-\infty}^\infty\hat{E}_s \mathcal{W}_{n-s}(R/L_{inj})^{1+m}~,
\label{eq:csf}
\end{align}
where we have used the fact that $\hat{\mathcal{A}}_p$ of KLP16 
is equal to $\widehat{E}_p$ of LP12 to change the notation to that
of Equation~(\ref{eq:Dn+}).

\subsubsection{The ratio of the multipole expansions of magnetic field angle and velocity centroid structure functions}

The amplitude of velocity perturbations $A_v$ has the same relation
to the variance of $\langle \delta V^2 \rangle$ as $A_B$ has
to $\langle \delta B^2 \rangle$. 
Thus ratios of the centroids and angle structure function
multipole coefficients can be expressed with the help of variances
as
\begin{equation}
\frac{\mathcal{D}_n}{D_n^\theta} =
\bar B_\perp^2 \frac{\left\langle \delta V^2 \right\rangle}{\left\langle \delta B^2 \right\rangle }
\times
\frac{\sum_{p}\hat{E}_p\mathcal{W}_{n-p}^{(A,F,S)}(\gamma)}
{\sum_{p} \hat{E}_p G_{n-p}^{(A,F,S)}(\gamma)}
\label{eq:Dc_over_Dtheta}
\end{equation}
We note that the residual dependence on the orientation
of the magnetic field with respect to the line of sight is arising
primarily due to different geometrical structure of 
the velocity and perpendicular magnetic fields as expressed
in distinct geometrical weights $\mathcal{W}(\gamma)$ and $G(\gamma)$.

\subsubsection{The effect of the composition of MHD modes}

Now we need to choose the composition of MHD turbulence in terms of energies in different modes.As we discussed in Appendix \ref{app:mhdturb} the velocity gradients of slow and Alfven modes trace magnetic field the same way, while the fast modes produce a perpendicular orientation of gradients. To have a discussion relevant to both to polarization and to gradients we defer considering the fast mode to the subsequent publication. This partly justified by the fact that fast modes do not dominate in MHD turbulence and they are subject to damping that is usually stronger than for Alfven modes. Therefore, dealing with turbulence at small scales we may frequently disregard the contribution of fast modes. For incompressible driving that we employ in our numerical simulation in order to test our expressions, fast modes are subdominant at all scales (see \citealt{CL02}). We shall discuss two of the simple cases here and work on the low $\beta$ case with numerical analysis instead (See \S \ref{sec:num}).
\linebreak

\noindent{\bf Pure Alfven case:} Note that purely Alfvenic case in our approximation has zero fluctuations if the mean field is perpendicular to the line of sight due to our formal
integration over the line of sight that set $k_z=0$. In LP12 we explained that in the mean magnetic field system of reference for finite turbulent Alfven Mach number $M_A$ one should account for magnetic field wandering which increases with $M_A$. This allows avoiding degeneracies that arise due to the excessively idealized setting.  However, some suppression of Alfvenic perturbation in line-of-sight projection for perpendicular field should be real effect. Accounting for the mean magnetic field changes along the line of sight leads to partial isotropization of the geometrical
effects in Equation~(\ref{eq:Dc_over_Dtheta}) which can be modeled by weighted addition of an isotropic term to geometrical functions as 
$G_{n-p}^{(A)} \to W_I(M_A) \delta_{np} + W_L(M_A) G_{n-p}^{(A)} $ and $\mathcal{W}_{n-p}^{(A)} \to W_I(M_A) \delta_{np} + W_L(M_A) \mathcal{W}_{n-p}^{(A)}$. Following
suggestion of LP12, one can adopt a simple model 
\footnote{This model
corresponds to assuming that at low Alfv\'enic Mach numbers,
the tangent of the typical deviation $\Delta\phi$
of the local direction of the magnetic field from the global mean one is given by $M_a$ and therefore $\overline{cos(\Delta\phi)^2} \approx 1/(1+M_A^2) $, 
while at large $M_a$ the field wandering angle covers all the values from 0 to $\pi/2$, thus $\overline{cos(\Delta\phi)^2} \approx 1/3$. We use this opportunity to note an inconsistency in LP12 where $\mathcal{W}_I$ as used
in Equation~(71) is twice the one introduced in  Equation~(45).}
\begin{equation}
\mathcal{W}_I \approx \frac{M_A^2}{1+ \sfrac{3}{2} M_A^2}, \quad
\mathcal{W}_L \approx \frac{1}{1+ \sfrac{3}{2} M_A^2}~.
\label{eq:WIWL}
\end{equation}
Retaining only the monopole $\hat{E}_0$ and quadrupole $\hat{E}_{\pm 2}$ in the power spectrum expansion, we obtain
\begin{equation}
\frac{\mathcal{D}_0}{D_0^\theta} \approx \aleph^{-1}
\bar B_\perp^2 
\frac{\mathcal{W}_I + \mathcal{W}_L (1-\cos\gamma (1 + \mathcal{K}_2(\gamma))) }
{\mathcal{W}_I + \mathcal{W}_L \cos\gamma (1 + \mathcal{K}_2(\gamma))}
\end{equation}
where 
\begin{equation}
\begin{aligned}
\mathcal{K}_2(\gamma) &= 2 \frac{\hat{E}_2}{\hat{E}_0} \frac{1-\cos\gamma}{1+\cos\gamma}\\
 \aleph &=\frac{\left\langle \delta B^2 \right\rangle}{ \left\langle \delta v^2 \right\rangle} \approx
4 \pi \bar{\rho}
\end{aligned}
\end{equation} Then, using
Eq.~(\ref{eq:WIWL}), one can obtain the strength of perpendicular magnetic field as
\begin{equation}
\bar{B}_\perp = \sqrt{4 \pi \rho}
\sqrt{\frac{\mathcal{D}_0}{D_0^\theta}}
\frac{1+M_A^{-2} \cos\gamma (1 + \mathcal{K}_2(\gamma))}
{1+M_A^{-2}[1-\cos\gamma (1 + \mathcal{K}_2(\gamma))] }
\label{B_alfen}
\end{equation}
\linebreak

{\noindent}{\bf High $\beta$ ($M_A<1$) case:} The situation is much simpler for the case of strong turbulence in high $\beta\propto M_A^2/M_s^2$ plasma, where both Alfvenic and slow modes are excited with the same power.  Physically this corresponds to the case of incompressible MHD turbulence. In this case $W_{p-n}^{A+S} = G_{n-p}^{A+S} = \delta_{np}$ so that we obtain
\begin{equation}
\frac{\mathcal{D}_n}{D_n^\theta} = \bar B_\perp^2 \aleph^{-1}
\end{equation}
Angular dependencies disappear despite the anisotropic
distribution of power in each mode, since the motion structure
is isotropic for such a mix. Same relation between variances of velocity and
magnetic field fluctuations is also true \footnote{We expect that with contributions of the fast modes, we shall have a "f" factor as we have in Eq.\ref{mean_B}.} for
the mix of Alfvenic and slow modes (high $\beta$) since
in this case these modes are two just polarization of the same motions and have equal power. We then conclude that
\begin{equation}
\bar{B}_\perp = \sqrt{4 \pi \rho}
\sqrt{\frac{\mathcal{D}_n}{D_n^\theta}},
\label{B_slow_alfen}
\end{equation}
which holds for all possible $n=0,2,4...$. Eq.\ref{B_slow_alfen} is a very simple expression for magnetic field strength and very similar to that given by Eq. (\ref{mean_B}). 
\linebreak

\subsection{Uncertainties and applicability}

Due to its simplicity Eq.(\ref{B_slow_alfen}) can be considered as our major result that we can recommend for the practical observational studies. The equal admixture of Alfven and slow modes is a good approximation to the weakly compressible MHD turbulence. To move further one requires to know a more detailed composition of MHD turbulence in terms of fundamental modes. This faces both theoretical and practical difficulties. On the practical side, the procedures of decomposition of contributions from different modes (see \citealt{KLP17a}) have not been applied to observations. On the theoretical side, the shocks formed by turbulent motions do not fit well into the picture of fast modes. All these issues deserve a rigorous study to be done elsewhere. We should just add here that qualitatively one expects to overestimate the strength of magnetic field if fast modes are present. 

As we mentioned earlier, Eq.(\ref{B_slow_alfen}) is formally very similar to the one obtained via our simplified approach in Eq. (\ref{mean_B}). The difference, however, that with our detailed approach we understand nature of the approximation that is used to obtain this expression. We also can see the nature of the uncertainties that are related to the practical use of Eq. (\ref{B_slow_alfen}). 

Incidentally,  Eq. (\ref{B_slow_alfen}) provides the estimate of magnetic field strength without the requirement of turbulence to have power law for all scales. By measuring the structure functions one localizes the contribution of the scales corresponding to the separation of the line of sight. Therefore it is enough to have the turbulence around this scale.  

Note, that Eq.(\ref{B_slow_alfen}) uses only the monopole part of the multipole decomposition in LP12. This monopole part can be easily obtained via isotropic averaging of observational data. In this paper we did not use the higher moments of the LP12 multipole decomposition, in particular, we did not use the quadropole term. This term carries the information about the anisotropy imposed by the on turbulence by the presence of the mean field. The amplitude of this term is another source of the information on the Alfven Mach number of turbulence. Naturally, this provides synergy and additional testing of the way of evaluating the magnetic field strength that we discuss in this paper. Making use of this quadropole term is the goal of our further studies. 

\begin{deluxetable}{c c c c c}
\tablecaption{\label{tab:sim} Description of MHD simulation cubes {  which some of them have been used in the series of papers about VGT \citep{YL17a,YL17b,LY18a,LY18b}}.  $M_s$ and $M_A$ are the R.M.S values at each the snapshots are taken. }
\tablehead{Model & $M_S$ & $M_A$ & $\beta=2M_A^2/M_S^2$ & Resolution }
\startdata \hline
huge-0                  & 6.17  & 0.22 & 0.0025 & $792^3$ \\
huge-1                  & 5.65  & 0.42 & 0.011 & $792^3$ \\
huge-2                  & 5.81  & 0.61 & 0.022 & $792^3$ \\
huge-3                  & 5.66  & 0.82 & 0.042 & $792^3$ \\
huge-4                  & 5.62  & 1.01 & 0.065 & $792^3$ \\
huge-5                  & 5.63  & 1.19 & 0.089 & $792^3$ \\
huge-6                  & 5.70  & 1.38 & 0.12 & $792^3$ \\
huge-7                  & 5.56  & 1.55 & 0.16 & $792^3$ \\
huge-8                  & 5.50  & 1.67 & 0.18 & $792^3$ \\
huge-9                  & 5.39  & 1.71 & 0.20 & $792^3$ \\ 
e6r3 (time-series)      & 5.45  & 0.24 & 0.0019 & $1200^3$\\\hline
Ms0.2Ma0.2 & 0.2 & 0.2 & 2 & $480^3$ \\
Ms0.4Ma0.2 & 0.4 & 0.2 & 0.5 & $480^3$ \\ 
Ms4.0Ma0.2 & 4.0 & 0.2 & 0.005 & $480^3$ \\
Ms20.0Ma0.2 & 20.0 & 0.2 & 0.0002 & $480^3$ \\ \hline
incompressible  & 0 & 0.7 & $\infty$ & $512^3$ \\ \hline\hline
\enddata
\end{deluxetable}

\section{Method}
\label{sec:method}

Most of the numerical data cubes {\re are} obtained by 3D MHD simulations that is from a single fluid, operator-split, staggered grid MHD Eulerian code ZEUS-MP/3D to set up a three dimensional, uniform, isothermal turbulent medium. To simulate the part of the interstellar cloud, periodic boundary conditions are applied. These simulations use the Fourier-space forced driving solenoidal driving.\footnote{Our choice of force stirring over the other popular choice, i.e. of the decaying turbulence, is preferable because only the former exhibits the full characteristics of turbulence statistics, e.g power law, turbulence anisotropy, extended from $k=2$ to {\re a dissipation scale of $12$ pixels}  in a simulation , and matches with what we see in observations (e.g. \citealt{1995ApJ...443..209A,2010ApJ...710..853C}). } For isothermal MHD simulation without gravity, the simulations are scale-free. If $V_{inj}$ is the injection velocity, while $V_A$ and $V_s$ are the Alfven and sonic velocities respectively, then the two parameters, namely, the Alfven Mach numbers $M_A=V_{inj}/V_A$ and sonic Mach numbers $M_s=V_{inj}/V_s$, determine all properties of the numerical cubes and the resultant simulation is universal in the inertial range. That means one can easily transform to any arbitrary units as long as the dimensionless parameters $M_A,M_s$ are not changed. The chosen $M_A$ and $M_s$ are listed in Table \ref{tab:sim}. For the case of $M_A<M_s$, it corresponds to the simulations of turbulent plasma with thermal pressure smaller than the magnetic pressure, i.e. plasma with $\beta/2=V_s^2/V_A^2<1$. In contrast, the case that is $M_A>M_s$ corresponds to the magnetic pressure dominated plasma with $\beta/2>1$. To investigate the behavior of the incompressible case, we adopt the incompressible cube our previous work \cite{Letal17}.

Further we refer to the simulations in Table \ref{tab:sim} by their model name. For example, the figures with model name indicate which data cube was used to plot the corresponding figure. Each simulation name follows the rule that is the name is with respect to the varied $M_s$ \& $M_A$ in ascending order of confinement coefficient $\beta$. The selected ranges of $M_s, M_A, \beta$ are determined by possible scenarios of astrophysical turbulence from subsonic to supersonic cases. 

\section{Numerical tests}
\label{sec:num}

\subsection{Building up the numerical recipe for incompressible MHD turbulence}
\label{subsec:recipe}

\begin{figure}[th]
\centering
\includegraphics[width=0.49\textwidth]{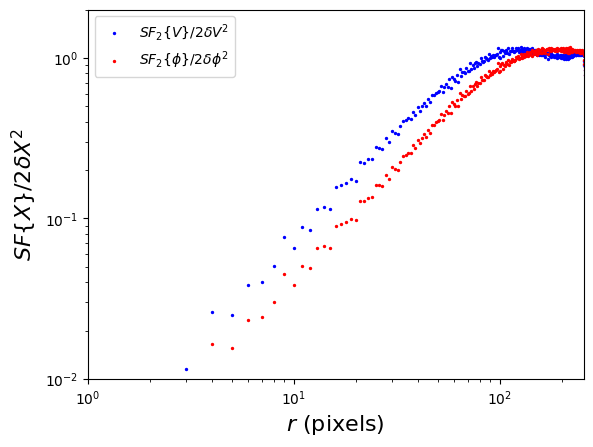}
\caption{\label{fig:illus1}  The angular averaged structure functions for the density-constant velocity centroid $V$ and the polarization angles $\phi$ normalized by $2$ times their variance respectively in an incompressible MHD simulation with $M_A=0.7$. }
\end{figure}

To use Eq.\ref{B_slow_alfen} practically, the two structure functions ${\mathcal{D}_n}$ and ${D_n^\theta}$ need to have the same power law with respect to distance, i.e. ${\mathcal{D}^V} \propto {D^\phi} \propto r^m$ for some $m$ {\it with distance $r$ smaller than the injection scale}. This requirement is easily fulfilled in the inertial range of incompressible sub-Alfvenic magnetized turbulence. For instance, Fig \ref{fig:illus1} shows the behavior of structure functions for both velocity and magnetic variables in 3D and projected 2D space in an incompressible magnetized turbulence with $M_A=0.7$. For easier visual comparisons we normalize the structure functions by the variance of the respective variables since $SF\{v\}(R\rightarrow\infty) \rightarrow 2 \langle \delta v^2 \rangle$.In Fig \ref{fig:illus1} we plot the angular average structure functions, i.e the monopole term of their angular dependence, for instance  for the velocity one
\begin{equation}
SF_{2D}\{V\}(R) = \frac{1}{2 \pi}\int d\theta SF_{2D}\{V\}({\bf R})
\end{equation}
The angular averaged structure functions are plotted in the distance range of $0,L/2$ where $L$ is the size of the simulation region, in our case $L=512$ pixels. Fig \ref{fig:illus1} shows the 2D structure functions for projected velocities $V = \int dz v$ and polarization angles $\phi=0.5\tan_2^{-1}(U/Q)$, in which they have the same power-law slope as a function of $r$ when $r\sim 20-60$ pixels. We shall utilize the range of scales that the two structure functions have the same power-law slope for the estimation of magnetic field strength.

Traditional DCF technique uses the ratio of $\delta v$ to $\delta \phi$ as an estimation of magnetic field strength (weighted by $\sqrt{4\pi\bar{\rho}}$. The use of the dispersions of $v$ and $\phi$ correspond to the part of their respective structure functions in Fig.\ref{fig:illus1} that has a {\it flat} slope and has $r\sim L_{inj}$. Hence the ratio of the dispersion functions, aka the structure functions with $r\ge L_{inj}$ would not be a function of distance. However the Alfven relation (Eq\ref{eq:alf}) develops only in scales smaller than the characteristic scales of the magnetized turbulence (See \S A) and these scales are smaller than $L_{inj}$. The fundamental physical issue for the DCF technique that utilizes the dispersion of observables in an unphysical length scale could be addressed properly by the structure function treatment which we are delivering below.

\begin{figure}[th]
\centering
\includegraphics[width=0.49\textwidth]{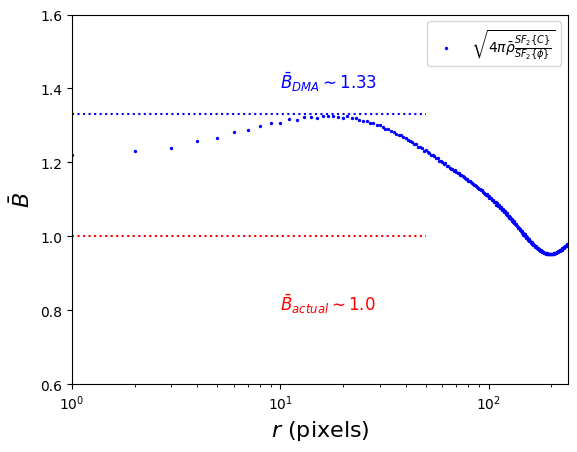}
\caption{\label{fig:sf2} The estimated magnetic field strength using DMA method as a function of distance by Eq.\ref{B_alfen} (blue). The exact value of $\bar{B}$ is drawn as a red dash line while the estimated value from the DMA method is marked with a blue dash line. }
\end{figure}

We shall seek for the part of the two structure functions $SF_{2}\{V\}({\bf R})$ and $SF_{2}\{\phi\}({\bf R})$ that have the same power-law slope. The length scale needs to be smaller than $L_{inj}$ and larger than the numerical dissipation scale, which is usually $12-16$ pixels depending on the properties of the numerical solvers. We shall use the upper bound of the numerical dissipation scales for our current analysis. As we show in Fig.\ref{fig:sf2}, the part of the two structure functions that carry the same power-law slope would be $r=20-60$ pixels. That means we could examine whether the quantity $\sqrt{4\pi\bar{\rho}SF_2\{C\}/SF_2\{\phi\}}$ would be approximately a constant of $r$ to obtain the magnetic field strength {\it at the length scales of $r=20-60$ pixels.} From Eq.\ref{fig:sf2}, the $\sqrt{4\pi\bar{\rho}SF_2\{C\}/SF_2\{\phi\}}$ is flat at in the length scale of $r \sim 20$ and giving $B_{estimated} \sim 1.3$, which is close to the global mean value of the magnetic field strength. We shall call the condition of obtaining magnetic field strength by comparing the ratio the structure functions of $SF_{2}\{C\}({\bf R})$ and $SF_{2}\{\phi\}({\bf R})$ that have the same power-law dependencies with respect to the distance $r$ be the {\bf flat criterion} for DMA. 

 The use of the structure functions for both velocities and magnetic field observables is advantageous compared to the dispersion method and also the Hildebrand-Houde method  since it provides a unique treatment of obtaining {\it local} magnetic field strength with less sampling points. For instance, one needs to compute only the structure functions with the distance lag $r\ge60$ pixels in our sample synthetic observations (Fig.\ref{fig:illus1}). This allows observers to acquire the magnetic field strength using smaller number of spectral information compared to the traditional DCF technique \citep{1951PhRv...81..890D,CF53}.

\subsection{Proceeding to compressible magnetized turbulence}
\label{subsec:compressible}

Using Eq.\ref{B_slow_alfen} in the case in compressible turbulence becomes more complicated because of the existence of the two compressible modes. \S \ref{subsec:detailed} discussed already how the combination of Alfven and slow modes would contribute to the structure functions and also the differential treatment in Eq.\ref{B_slow_alfen}. { Indeed, in the presence of the compressible modes, the structure functions of velocities and polarization angles are expected to behave differently from what we see from the incompressible counterpart since the slope of structure functions are closely related to the slope of the power spectrum, and the fast modes have different power spectral slopes ($P_{F}(k)\propto k^{-3/2}$) than that of Alfven and slow modes ($P_{A,S}(k_\perp,k_\parallel \sim k_\perp^{2/3})\propto k_{\perp}^{-5/3}$) even in small $M_s$ case (See \citealt{CL03}). Therefore the recipe that we developed in \S \ref{subsec:recipe} would not work unless we have an adjustment on the case when the two structure functions $SF_{2}\{C\}({\bf R})$ and $SF_{2}\{\phi\}({\bf R})$ have different power-law slope. 
}

\begin{figure}[th]
\centering
\includegraphics[width=0.48\textwidth]{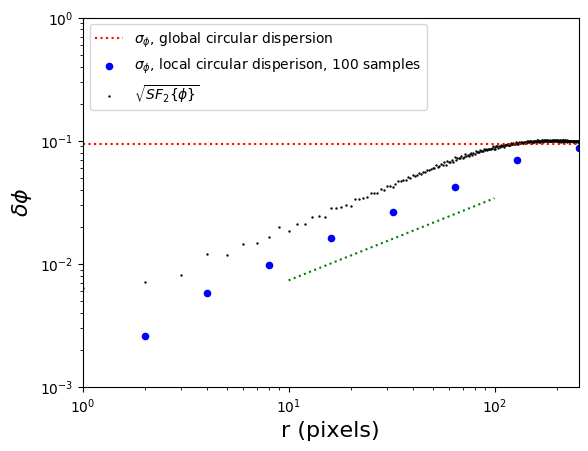}
\caption{\label{fig:picking_correct_scale} A figure showing how the structure function of polarization angle is related to the its circular dispersion when sampled locally. We first randomly select a square block of size $r$ (pixels) and compute the circular dispersion within it. The averaged value of 100 such selections are plotted as the "local circular dispersion" value as the blue points in this figure as a function of block size $r$. For reader's comparison, we also plot the square root of the angular averaged structure function (black curve) and the global circular dispersion (red dash line).}
\end{figure}

\begin{figure*}[th]
\centering
\includegraphics[width=0.48\textwidth]{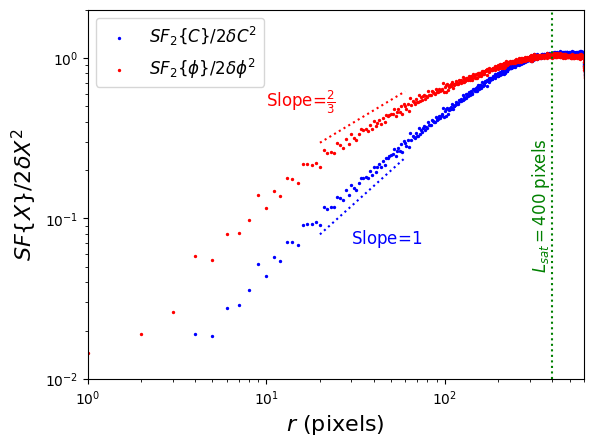}
\includegraphics[width=0.48\textwidth]{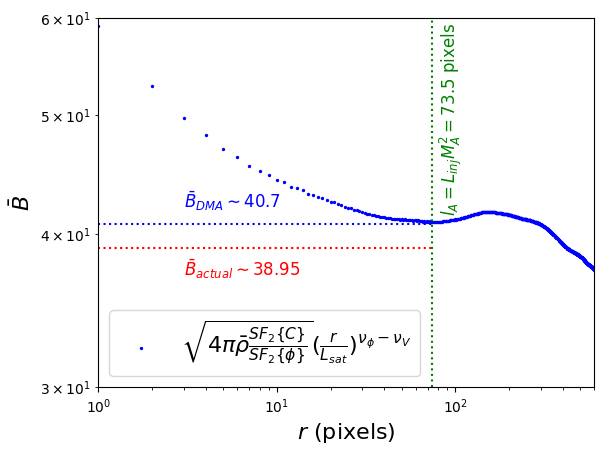}
\caption{\label{fig:compressible} (Left) The structure functions of the normalized velocity centroid $C$ (blue) and the polarization angle $\phi$ (red). The respective trend lines ($\nu_\phi = 2/3, \nu_v=1$) are added at the distance range of $r=20-60$ pixels. (Right) A plot showing the value computed by Eq.\ref{eq:compensated_B} as a function of distance $r$ (blue scatter points). We search for the part of the curve that has a flat slope, which is indicated by the blue dash line. The inferred magnetic field strength by Eq.\ref{eq:compensated_B} is close to the exact value which is indicated with the red dash line.  }
\end{figure*}

We use the method of {\it compensated structure functions} as a workaround for using Eq.\ref{B_alfen} or Eq.\ref{B_slow_alfen} and extends the latter equations to local structure functions and also local dispersions. The idea is illustrated in Fig.~\ref{fig:picking_correct_scale} with the introduction of the local statistical quantities such as the local circular dispersions. The local circular dispersion is simply an extreme case of the structure function since one should recall $SF(R\rightarrow L) = 2\sigma^2$ if $L$ is the size of the region. 
For reader's comparison, we also plot the square root of the angular averaged structure function (black curve of Fig.~\ref{fig:picking_correct_scale}) and the global circular dispersion (red dash line of Fig.~\ref{fig:picking_correct_scale}). One could see that the local circular dispersion actually follows the same power-law as the square-root of the angular averaged structure function. This implies that if we are limited to a small area for sampling, assume its size is $r^2$ while the characteristic scale for the cloud to be $L_{cloud}$, then we can estimate locally the dispersion of angles by 
\begin{equation}
    \delta \phi_{local} = \sigma_{\phi,local}\left(\frac{L_{cloud}}{r}\right)^{\nu_\phi/2}
\end{equation}
The respective CF method is, formally: 
\begin{equation}
B \sim \sqrt{4\pi\bar\rho} \frac{\delta v}{\sigma_{\phi,local}}\left(\frac{L_{cloud}}{r}\right)^{-\nu_\phi/2}
\end{equation}
The formula should subject to the theoretical correction in \S \ref{subsec:detailed}. We also expect if we take the total differential measure approach , then both $\sigma_v$ and $\sigma_\theta$ would have a distance compensation factor of $\left(\frac{L_{cloud}}{r}\right)^{\nu_\phi/2}$ for some structure function power indices $\nu_{V,\phi}$, which accounts for the insufficient statistical sampling on the sky. To utilize Eq.\ref{B_alfen}, the respective CF method should be 
\begin{equation}
B \sim \sqrt{4\pi\bar\rho\frac{SF_V}{SF_\phi}}\left(\frac{L_{cloud}}{r}\right)^{(\nu_V-\nu_\phi)/2}
\label{eq:compensated_B}
\end{equation}
If accidentally $\nu_V-\nu_\phi=0$ (most likely in sub-sonic,sub-Alfvenic or incompressible case), then there is no compensation term needed. Fig. \ref{fig:compressible} shows an example on how to utilize Eq.\ref{eq:compensated_B} when the structure functions of the two observables have different power-law dependencies with respect to $r$ in a compressible magnetized turbulence. The compressible simulation ``e6r3'' used here is a $1200^3$ super-sonic ($M_S=5.45$) sub-Alfvenic ($M_A=0.35$) saturated turbulence simulation, which as a plasma $\beta\ll 1$.In this scenario the wave-vector of the slow mode is expect to be parallel to the local mean magnetic field direction \citep{CL03}. 

We shall apply the flat criterion for the compensated structure functions.{ Notice that there is a particular length scale $l_A=L_{inj}M_A^2\sim 73.5$ pixels here for the Alfven relation (Eq.\ref{eq:alf}) to develop (See Appendix). We are therefore seeking for the flat criterion to hold with length scales $r\le l_A$}. Here we assume that we have the knowledge of $L_{inj} = L_{sat} = 400$ pixels (green dash line of Fig.\ref{fig:compressible}) where $L_{sat}$ is the length scale for the structure functions $SF_{2}\{C\}({\bf R})$ and $SF_{2}\{\phi\}({\bf R})$ to be saturated. Using the flat criterion as delivered in \S \ref{subsec:recipe}, we see that the DMA method with the length scale correction (Eq.\ref{eq:compensated_B}) has a very nice estimation of magnetic field strength (40.7) compared to the actual value (38.5), despite that we do not have the information of ratio of MHD modes on hand.

\subsection{Dependencies on $M_s$, $M_A$}
\label{subsec:fcond}

\begin{figure*}[th]
\centering
\includegraphics[width=0.48\textwidth]{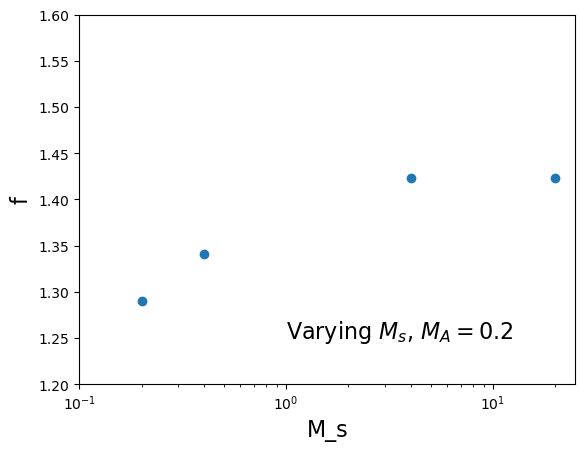}
\includegraphics[width=0.48\textwidth]{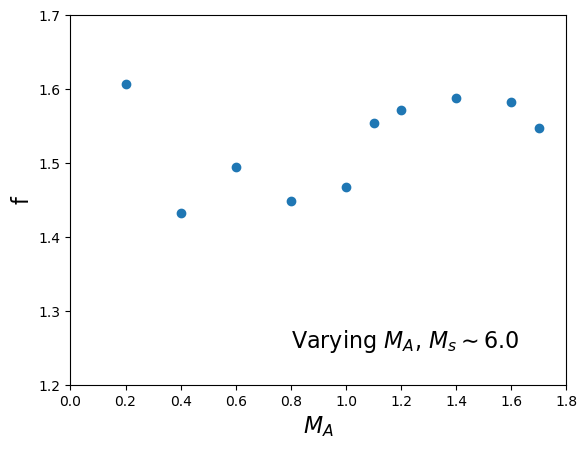}
\caption{\label{fig:f2} The response of the constant C as defined in Eq.\ref{mean_B} (c.f. Eq.\ref{centroid}) as a function of sonic and Alfvenic Mach number. For the group with varying sonic Mach number, their $M_A \sim 0.2$. While for that of varying Alfvenic Mach number, their $M_s$ are $\sim 6.0$. }
\end{figure*}

To apply Eq.~(\ref{mean_B}) or Eq.~(\ref{B_slow_alfen}) in observations, we need to know how the constant $f$ is related to the global properties of MHD turbulence (i.e. $M_s,M_A$). Knowing how the conversion factor is related to the sonic and Alfvenic Mach number is crucial for the DMA technique. Here we use the density weighted centroid (c.f. Eq.\ref{centroid}) as the density is not constant anymore in compressible turbulence for our testing of Eq.\ref{mean_B}. Fig. \ref{fig:f2} shows how the conversion factor is related to the sonic Mach number $M_s$ (left) and Alfvenic mach number $M_A$. One could see that while there is a tiny fluctuation on the value of $f$ as a function of $M_s$ and $M_A$, the fraction of fluctuation is relatively small ($\sim 10-20\%$) compared to the mean value. Therefore we conclude that we can take a range of value of $f \sim 1.3-1.6$ in observation.

\section{Use of multi-point statistics and suppression of the effects of to shear and self-gravity}
\label{sec:multipoint}

We have used for our study two point second order structure functions. Compared to correlation functions, those allow removing the constant shifts of the foreground. In the presence of shear we provided the procedure for removing the shear contribution. However, there is a more robust way of dealing with the problem that was explored for the statistical studies of emission lines in \cite{LP08}, namely, the use of multi-point structure functions. A detailed description of three and four point second order structure functions is given in \citeauthor{CL09} (\citeyear{CL09}, see also \citealt{2007PhRvL..98o4501F,LP08,Cho19}). 

For our approach the number of points does not matter, as the magnetic field strength enters the expression via the Alfvenic relation between the perturbations of magnetic field and velocity. Therefore with the multi-point structure functions we can use Eq. \ref{B_alfen} or \ref{B_slow_alfen} to determine the magnetic field strength.

\subsection{Theoretical description of DMA in the presence of galactic shear and regular velocity components}
\label{subsec:shear}

One of the advantage of the DMA is the ability of tackling regular shear flows through treatments of the structure functions on velocity observables. In this subsection we illustrate how to tackle this self-consistently. Let us suppose that a regular velocity field $v^{(rg)}$ is added on top of the 3D turbulent velocity $v^{(0)}$, so that the total
velocity is
\begin{equation}
v^{(t)}_i ({\bf r}) =   v^{(0)}_i({\bf r})   + v^{(rg)}_i({\bf r}) 
\end{equation}

If we approximate the regular velocities to linear order in expansion around the center $\mathbf{r}^0$ of the emitting volume, the effect of the regular 
motions
\begin{equation}
v^{(t)}_i ({\bf r}) =   v^{(0)}_i({\bf r})   + v^{(rg)}(\mathbf{r^0}) + v^{(rg)}_{ij}(r_j - r^0_j)    
\end{equation}
is determined by the linear shear tensor
\begin{equation}
    v^{(rg)}_{i,j} = \frac{\partial v^{(rg)}_i(\mathbf{r}^0)}{\partial r_j}
    \label{eq:tensor_expression}
\end{equation} 
Here Einstein summation over repeated indices is used and $i,j=x,y,z$.

Assuming incompressible turbulence at constant density, the velocity centroid $C({\bf R}_1) = \int dz  v_z^{(t)}({\bf R}_1,z)$ is the integral along the line of sight of total velocity $z$-component. Taking the difference
of the values of two centroids at sky separation $\mathbf{R}$ 
\begin{equation}
C({\bf R}_1+{\bf R}) - C({\bf R}_1) = \int dz v_z^{(0)}(\mathbf{R},z)
+ \mathcal{L} v^{(rg)}_{zl} R_l
\end{equation}
where $\mathcal{L}$ is the depth of the emitting volume and $l=x,y$, 
eliminates all terms in the regular velocity contribution that are constant across the sky. 

The standard 2-point structure function of the
centroids is obtained by averaging of the square of this difference
and is easily shown to be given by
\begin{equation}
\begin{aligned}
SF & \{C\}({\bf R}) \equiv \langle(C({\bf R}_1+{\bf R}) - C({\bf R}_1)]^2\rangle \\
& = {\cal L} \int dz  \Big( SF\{v^{(0)}\}({\bf R},z) - SF\{v^{(0)}\}({\bf 0},z) \Big)\\
& + {\cal L}^2 S^{(rg)}_{lm}  R_l R_m 
\end{aligned}
\end{equation}
where first term is the standard structure function of turbulent centroids, which behaves as $R^{1+m}$ if $SF\{v\} \sim r^m$. and the second term is due to regular shearing velocities with $S^{(rg)}_{lm} = v^{(rg)}_{zl} v^{(rg)}_{zm} $. We see that the turbulent term accrued one extra power of $R$ in projection,  while the regular term remains quadratic in $R$. Thus, a general model for the angle average two-point structure function of centroids that is consistent with MHD theory in the presence of regular shear has the 
form
\begin{equation}
SF\{C\}(R) = q R^{1+m} + p R^2 
\label{structure_f}
\end{equation}
where $q$ and $p$ are constants. For a realistic turbulence the first term saturates at the energy injection scale $L_{inj}$. In below we discuss how the multipoint structure functions would aid removing the shear contribution from the centroid statistics.

\subsection{Application of 3 and 4 point statistics to regions with shear}
The three \& four point second order structure functions are defined as :
\begin{equation}
\begin{aligned}
    &SF_{3pt}\{A\}({\bf r}) \propto\langle [A({\bf x}+{\bf r})-2A({\bf x})+A({\bf x}-{\bf r})]^2\rangle \\
    &SF_{4pt}\{A\}({\bf r}) \\&\propto\langle [A({\bf x}+2{\bf r})-3A({\bf x}+{\bf r})+3A({\bf x})-A({\bf x}-{\bf r})]^2\rangle ~.
    \label{eq:multiptSF2}
\end{aligned}
\end{equation}
 Their ability to remove shear velocity field and recover the original structure functions was illustrated in \cite{Cho19}. In a similar manner to 2-point structure function that cancelled constant velocity contribution,, 3-point and higher order structure functions cancel any linear and respectively higher power regular contributions to velocity field. The cost of using them is an increased noise contribution when applied to noisy data.
 
\begin{figure}[th]
\centering
\includegraphics[width=0.48\textwidth]{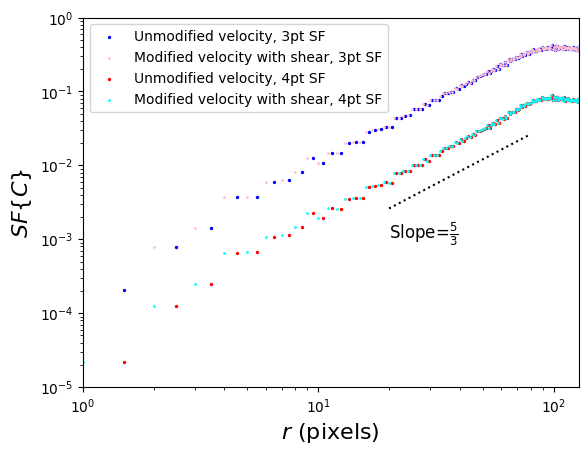}
\caption{\label{fig:sf_multipoint} A figure showing the behavior of the 3 point and 4 point projected velocity structure function with and without the presence of shear in the incompressible cube. To display the differences we introduce a small horizontal offset of the structure functions of unmodified velocity field, and also a vertical offset of 3 for both the 4 point structure function. }
\end{figure}

Here we test whether this is the case. Fig.~\ref{fig:sf_multipoint} shows how the 3-point and 4-point {\it angular averaged} structure functions behave as a function of the correlation lag $r$ on the projected velocities of incompressible cube. Notice that due to the non-periodicity of the map, using more points within a region decreases the possible number of statistical samples. We can see from Fig \ref{fig:sf_multipoint} that the multi-point structure functions are indeed not altered by the presence of constant velocity shear field. This could potentially replace the method used in \S \ref{subsec:shear} with a cost of reducing ranges of the correlation lag $r$.

\subsection{Applications of DMA to self-gravitating media}

One of the issue of the DCF technique is its questionable applicability in self-gravitating regions, through it is extensively applied to observations. Therefore we would like to examine whether the DMA equations (Eq.\ref{B_alfen} or Eq.\ref{B_slow_alfen}) need to be modified in the presence of strongly gravitating regions. 

Assuming we are given a vector field of polarization angle $\hat{\phi}$ defined at a space $A\in \mathcal{R}^2$, then the "polarization angle streamlines", which resembles the geometry of the magnetic field lines (See \citealt{curvature} for a detailed discussion). Following \cite{curvature},the definition of unsigned polarization angle curvature $\kappa$ would be
\begin{equation}
    \kappa =\left|\frac{d\hat{\phi}}{dl} \right|
    \label{eq:curvature}
\end{equation}
where $dl$ is the line element of the "polarization angle streamlines". In the case of structure functions, the effect of curvature would not accumulate until $\kappa r\sim 1$. Therefore the structure functions would have different power-law dependence when $r<1/\kappa$ and $r>1/\kappa$. To study this effect, we select a $200$ pixels $\times 200$ pixel region from a $(1200\text{pixels})^2$ synthetic observation map from "e6r3" (See Table \ref{tab:sim}) to investigate the effect the magnetic field line curvature to the structure functions.

\begin{figure*}[th]
\centering
\includegraphics[width=0.48\textwidth]{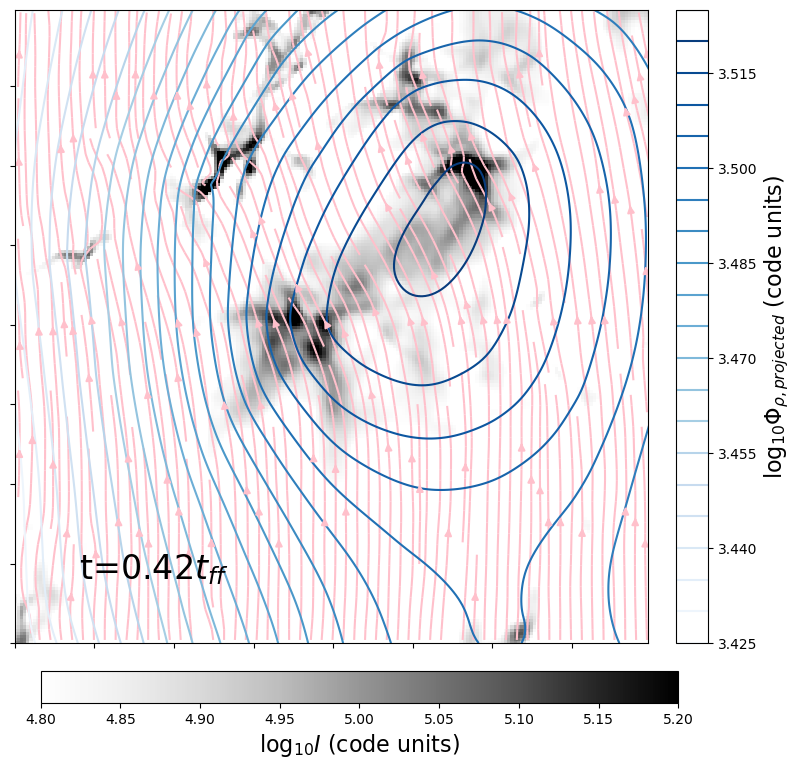}
\includegraphics[width=0.48\textwidth]{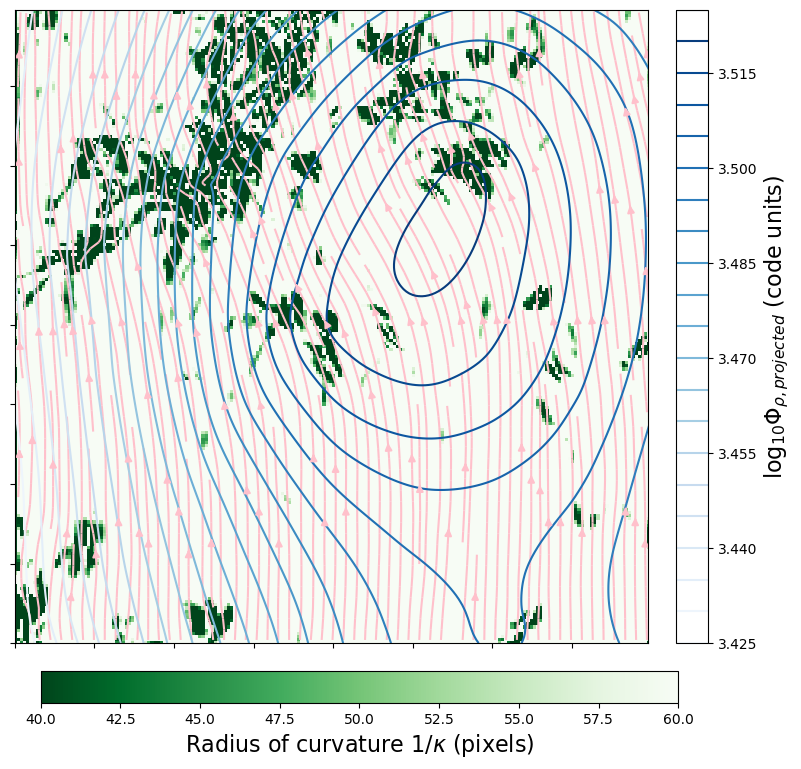}
\caption{\label{fig:grav} (Left) An image showing the intensity map (the Grey back-image) self-gravitating cloud from synthetic observations of a late stage of the time series of e6r3  with its magnetic field streamlines mapped by polarization angles plot as pink while the projected gravitational potential drawn with contours as blue. (Right) A map showing the radius of curvature $1/\kappa$ in the same area with color bar adjusted to $40-60$ pixels, showing the region that has strong curvature as dark green.}
\end{figure*}

For instance, the left of Fig. \ref{fig:grav} we show a self-gravitating simulation with its magnetic field streamlines (pink) and gravitational potential contours (blue). The simulation's snapshot is taken at $t=0.42t_{ff}$ where $t_{ff}$ is the free fall time. This particular snapshot is taken right before the Truelove criterion is violated (See \citealt{1997ApJ...489L.179T}). The curvature distribution is displayed in the right of Fig. \ref{fig:grav} which shows a significant area of the synthetically observed regions has the radius of curvature $1/\kappa<40$ pixels.

\begin{figure}[th]
\centering
\includegraphics[width=0.48\textwidth]{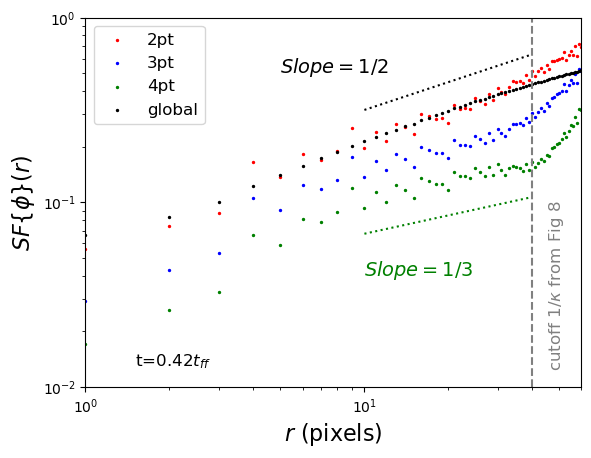}
\caption{\label{fig:grav2} A figure showing how the 2-point (red,2pt), 3-point (blue,3pt) and 4-point (green,4pt) structure functions behave as a function of $r$ compared to the case when we compute the global structure function (black,global) which contains zero curvature in average. The turning point suggests that there is a large scale magnetic field curvature with the radius of curvature of $r\sim 40$ pixels contributing to the dispersion of structure functions. Two trend lines are added to show the slope differences between the 3/4 point structure functions (green dash line, 1/3) and the global/2-point structure function (black-dash line). A vertical dash line marking $r=40$ is drawn to signify the effect of radius of curvature of $40$ pixels we have seen in Fig.\ref{fig:grav}. Vertical offsets are introduced for reader's easy visual comparisons between different structure functions. }
\end{figure}

We are interested to see how the change of number of points in computing structure functions would change the behavior of the structure functions as a function of distance. Fig. \ref{fig:grav2} shows the 2-point (red), 3-point (blue) and 4-point structure functions (green) computed in the area of interest (the $(200\text{pixel})^2$) region and also the 2-point structure function computed globally (black). We can see that all variants of the structure functions (2-point, 3-point, 4-point) are generally linear in log-log space until $r=40$ pixels, which can be visually seen by comparing the structure functions to that of the global structure functions computed in the $(1200 \text{pixels})^2$ area (black). The special scale $r=40$ pixels represent the bending that we can visually see on the right of Fig.\ref{fig:grav}. In fact, for the 3-point and 4-point structure functions there is a significant change of the slope of the structure functions that can hardly be seen in in the 2-point structure functions. This suggests that one potentially separate the large scale curvature contribution to that of the small scale fluctuation of magnetic field if one compares the local multipoint structure functions to the global structure functions, especially when applying the DMA to the self-gravitating regions.

\section{Comparison DMA with the earlier use of structure functions}
\label{sec:comparison}

In \cite{2009ApJ...696..567H} the structure functions of fluctuations of the polarization angle directions introduced in \cite{Fal08} were used in order to find magnetic field strength of molecular clouds. The model within \cite{2009ApJ...696..567H} which the results were obtained assumes that the correlation scale of magnetic turbulence is smaller than the lag $l$ between the points for which the structure function is calculated. Such calculations are applicable to very low resolution studies, e.g. the studies applicable to magnetic field in other galaxies. 

In Milky Way molecular clouds the turbulence spectrum can be resolved in molecular clouds. For instance, \cite{2011ApJ...733..109H} found spectra of turbulence $k^{-\alpha}$ with $\alpha = 1.4\pm 0.4$. Note, that the Kolmogorov value of $5/3$ is getting within this range. As for evaluating the strength of \cite{2011ApJ...733..109H}) adopted the traditional DCF technique and did not make use of the advantages of local measured provided by structure functions that we employ here.

The idea of using differential measures in estimating magnetic field properties is not new and has been explored by the community in different ways. \cite{EL05} uses the anisotropy of spectroscopic observables, e.g. velocity centroids, to estimate the orientation In the work of of magnetic fields, while \cite{2009ApJ...696..567H} and the subsequent publications \citep{2009ApJ...706.1504H,2011ApJ...733..109H,2012ApJ...749...45C,2016ApJ...820...38H} investigate a model of structure functions of observed polarization angles to estimate the the turbulent-to-regular magnetic field strength ratio and thus the strength of the regular part of magnetic field. In a separate development, \cite{LP12,LP16} discusses the properties of the structure function of synchrotron observables (intensity and polarization) bases on the theory of MHD turbulence. The analysis framework of \cite{LP12,LP16} has also been used to the studies of velocity centroid structure functions \citep{KLP17a}.

In the work of \cite{2009ApJ...696..567H}, they only replace $\delta \theta \rightarrow SF^{1/2}_{2D}\{\phi\}$ and investigate its properties under several important assumptions that lead to the estimation of $\langle B^2\rangle $: (1) The turbulence is isotropic; (2) There exist two length scales : the turbulent correlation scale $\delta$ and the large scale magnetic field scale $L$;  (3) The magnetic field can be written as the sum of the regular part and turbulent part ${\bf B} = \langle B\rangle + B_t$ with a special properties that $\langle B_t({\bf r'})B_t({\bf r}+{\bf r'})\rangle_{\bf r'}  = 0 $ for all $r\ge \delta$; (4) The two dimensional polarization angle (the observable of magnetic field angle in 2D) can be modelled by the Taylor expansions of structure functions of polarization angles:
\begin{equation}
    SF_{2,2D}\{\phi\}({\bf R}) \sim b^2 + m^2 R^2
    \label{eq:HH1}
\end{equation}
where $b,m$ are some fitting factors. It is shown in \cite{2009ApJ...696..567H} that the turbulent to regular magnetic field strength ratio to be:
\begin{equation}
    \frac{\langle B_t^2\rangle}{\langle B^2\rangle} \sim \frac{b^2}{2-b^2}
    \label{eq:HH2}
\end{equation}
and thus combining the DCF method,  \ref{eq:HH1},\ref{eq:HH2} and writing $SF_{2,2D}\{V\}({\bf R}) \rightarrow \delta v^2, \langle B_t^2\rangle/\langle B^2\rangle \rightarrow \delta \phi^2$, , \cite{2009ApJ...696..567H} arrives
\begin{equation}
    \langle B^2\rangle \sim (2-b^2)4\pi \langle \rho \rangle \frac{\delta v^2}{b^2}
    \label{eq:c-HH}
\end{equation}

\cite{2009ApJ...706.1504H} further expands the method developed in \cite{2009ApJ...696..567H} by considering the telescope beam effect and also introduces a Gaussian model for both the auto-correlation function and the beam profile function. Under such formalism, not only could they obtain the turbulent to regular magnetic field strength ratio but also the number of turbulent eddies $N$ (which they call "independent turbulent cells") along the line of sight which could be obtained by fitting the structure function similar to that in \cite{2009ApJ...696..567H}. It's worth to note that \cite{CY16} argues separately $\sqrt{N} \sim \delta C/\delta v_{los}$. Similar idea is behind our derivation in  Eq.(14).

The approach described in  \citep{2009ApJ...696..567H,2009ApJ...706.1504H} was applied to observational data to both molecular clouds \citep{2011ApJ...733..109H,2012ApJ...749...45C,2016ApJ...820...38H} and also galactic disks \citep{2013ApJ...766...49H}. However, we claim that the assumptions made within this approach are inconsistent to the theory of MHD turbulence (See \citealt{GS95,LV99,CL02,Cho19}) and the subsequent theoretical studies of properties of structure functions that arise from MHD turbulence (see \citealt{LP12,LP16,KLP17a}, \S \ref{sec:theory}). For instance, Eq. (\ref{structure_f}) presents what sort of structure functions we expect to see in turbulence in the presence of shear.
In view of that, below we propose an alternative explanation of some of the observational data.

As we discuss in Appendix A2, if turbulence is super-Alfvenic, i.e. $M_A>1$, the magnetic fields are correlated up to the scale $l_A$ given by Eq. (\ref{la}), i.e. $l_A=L_{inj}M_A^{-3}$. This scale can be found by correlating the the structure functions of polarization angle directions, for instance (see \citealt{Fal08,2009ApJ...696..567H}). The turbulent injection scale $L_{inj}$ can be found by correlating velocity  fluctuations. A more sophisticated ways of measuring $L_{inj}$ are also possible (see  \cite{2010ApJ...710..853C}). With some additional assumptions one can identify $L_{inj}$ by the analysis of density fluctuations.  

Some observational results were interpreted within Hildebrand-Houde approach. Below we provide an alternative interpretation of the data. For instance,
in \cite{2012ApJ...749...45C} it was claimed that magnetic field fluctuations in OMC-1  have the correlation scales $\sim 9^{"}$ and $\sim 7^{"}$ for the Stokes Q and U parameters and $\sim 13^{"}$ for density fluctuations. From the theory of super-Alfvenic turbulence (see Appendix A2), the first two numbers can be associated with the angular size associated with $l_A$, and the third number with the injection size $L_{inj}$. Using Eq. (\ref{la}) one can estimate $M_A$ for OMC-1 as $(13/7)^{1/3}\approx 1.2$. This means that OMC-1 is a mildly super-Alfvenic object.

\section{Magnetic Field strength from combining $M_s$ and $M_A$}
\label{sec:VGTDMA}

\subsection{Application to channel maps}
The method that we developed in \S \ref{sec:theory} can also be resembled with the turbulent Mach numbers that we could obtain from different techniques. From \S \ref{subsec:detailed} we see that it might be possible that there are extra weighting factors as a function of $M_A$ depending on the mode composition. In the current section we shall stick with the simplest form of the DMA, i.e. the DCF technique form Eq.\ref{eq:BB} for our discussion.

As we discussed in \S \ref{sec:strength}, in observations the measure that is directly available with polarization measurement is $M_{A, \bot}=\delta v/V_{A,\bot}$, where $\delta v$ is the injection velocity and $V_{A,\bot}=B_{\bot}/\sqrt{4\pi\bar{\rho}}$ is the plane of sky Alfven velocity. The same value can be presented as  $M_{A, \bot}=\delta B_{\bot}/B_{\bot}$, where $\bot$ denotes the plane of sky magnetic field component. 

Notice that the value of the perpendicular component  the magnetic field can be obtained from the ratio 
\begin{equation}
    \frac{M_s}{M_{A,\bot}}=\frac{B_{\bot}}{c_s} (4\pi \rho)^{-1/2}.
\end{equation}
Writing $\Omega=\delta v_{\parallel}/\delta v$ and noticing that $M_A \sim \delta v/v_{A,\bot}$, we would have
\begin{equation}
    B_{\bot}={\Omega}c_s\sqrt{4\pi \rho}M_s M_{A,\bot}^{-1}
    \label{eq:channelb}
\end{equation}
The emergence of the geometric term $\Omega \delta v_{\parallel}/\delta v$ suggests that there is a geometrical factor that affects the measurable magnetic field strength on the plane of sky.  For Alfvenic waves perpendicular and magnetic field perpendicular to the line of sight $\Omega\approx 1$. In general, it depends on the nature of turbulence and the angle $\gamma$ given by $\Omega \sim \sin\gamma$. However, the geometrical factor is not trivial since it is a complex function of the Alfvenic Mach number. We shall discuss how these geometrical factor would affect the measurement in a later work by \cite{BLOS}. In the time being, we shall study the case when $\Omega=1$, i.e. $B\perp LOS$.  The technique that we introduce here uses two different Mach numbers, the Alfven and sonic one. Therefore we will term this technique MM2.

One special property about the use of Eq.\ref{eq:channelb} is that, both $M_s$ and $M_A$ can be obtained purely from spectroscopic channel map and $c_s$ can be obtained from the accepted measurements of temperature of emitting gas \citep{2006ApJ...636.1114D} or by the thermal deconvolution method described in\citep{2018arXiv180200024Y}. As for the value of $M_A$, it can be obtained both polarization and non-polarization method, e.g. using the width of the probability distribution function of the gradient directions within a sub-block (see \citealt{2018ApJ...865...46L}) or the curvature of either polarization or gradient orientations (see \citealt{curvature}).  As a result the value of $M_{A,\bot}$ can be obtained in a more localized fashion compared to the observations of polarization. The most striking advantage of MM2 is that applying velocity gradients to channel maps it is possible to find the distribution of $M_{A,\bot}$ for channel maps. The corresponding distributions of $M_{A,\bot}$ were obtained with channel maps for galactic HI in \cite{2018ApJ...865...46L} and for molecular CO lines in \cite{velac}.  

The sonic Mach number $M_s = \delta v/c_s$, where $c_s$ is the sound velocity, can also be obtained using the statistical properties of the velocity channel maps. Various techniques of obtaining $M_s$ are suggested. For instance,  \cite{2010ApJ...708.1204B} \& \cite{BL12} successfully used the skewness and kurtosis of the intensity PDFs and established the relation between these quantities and $M_s$. This technique was successfully applied it to HI in Small Magellanic Cloud to find the POS distribution of $M_s$.

Other ways of obtaining $M_s$ include the analysis of Tsallis statistics \citep{2011ApJ...736...60T} and a more recent technique based on using the distribution of amplitudes of velocity gradients \citep{2018arXiv180200024Y}.  Similar to PDFs, the calculation of $M_A$ with velocity gradients can be done locally, as it is demonstrated in \cite{2018ApJ...865...46L}. The results of $M_s$ measurements are very robust and they are marginally affected either shear or by large-scale magnetic field curvature. Similarly, the calculation of $M_s$ is not much influenced by the galactic shear or any other large-scale shear induced by non-turbulent motions. Therefore, like the DMA, the MM2 technique is local and, compared to the DCF, it can be used for a wider variety of astrophysical settings.

\begin{figure}[th]
\centering
\includegraphics[width=0.48\textwidth]{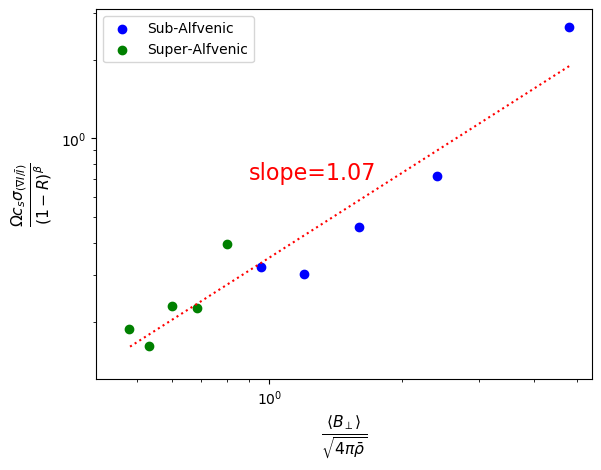}
\caption{\label{fig:channelc} A figure showing how the quantity $\Omega c_s M_s/M_A \sim c_s\sigma_{\left(\frac{\nabla I}{\bar{I}}\right)}(1-R)^{\beta}$ should be related to the Alfven speed $v_A=\langle B\rangle/\sqrt{4\pi\bar{\rho}}$ based in Eq.\ref{eq:channelc} in the set of simulations huge-0 to huge-9. We draw the points that are sub-Alfvenic as blue while those who are super-Alfvenic as green. }
\end{figure}

In the regime of the Velocity Gradient Technique, the ratio of gradient amplitude to gradient dispersion provides the expression $M_s/M_{A,\bot}$. In particular, the gradient observables could be related to the Mach numbers :
\begin{equation}
\begin{aligned}
    \sigma_{\left(\frac{\nabla I}{\bar{I}}\right)} &\propto
    \begin{cases} 
    M_{s}^{2} \quad &(M_s<1)\\
    M_{s} \quad &(M_s>1)
    \end{cases}\\
    1-R &\propto
    \begin{cases} 
    M_{A,\perp}^{-0.14\pm0.03} \quad &(M_{A,\perp}<1)\\
    M_{A,\perp}^{-0.06\pm0.03} \quad &(M_{A,\perp}>1)
    \end{cases}
\end{aligned}
\label{eq:GAGD}
\end{equation}
where $\nabla I/\bar{I}$ is the gradient amplitude of normalized intensity \citep{2018arXiv180200024Y} and $1-R$ is the inverted variance for twice of the gradient angle orientation \citep{2018ApJ...865...46L}\footnote{The circular standard deviation of VGT is defined as $\sigma_{VGT} = \sqrt{-2\ln(R)}$ if R is known, i.e. $1-R = 1-e^{-\sigma^2_{VGT}/2}$. When $\sigma_{VGT}$ is small, $1-R\sim\sigma^2_{VGT}/2$}. This approach requires to use the unique properties of velocity gradients, namely, that the same volume of emitting gas is used both to find $M_s$ and $M_{A,\bot}$. As a result, using the Galactic rotation curve one can obtain the distribution of the value of the plane of sky component of galactic magnetic field at different distances from the observer. In fact, we can approximate Eq.\ref{eq:channelb} into the combinations of gradient amplitude and gradient dispersion assuming $M_s>1$:
\begin{equation}
    B_{\bot}\propto C{\Omega}c_s\sqrt{4\pi\bar{\rho}}\sigma_{\left(\frac{\nabla I}{\bar{I}}\right)}(1-R)^{\beta} \quad (M_s>1)
    \label{eq:channelc}
\end{equation}
where the constant $C$ contains the proportionality constants related to the techniques in \cite{2018arXiv180200024Y,2018ApJ...865...46L}. Moreover, $\beta=1/0.14$ for $M_A<1$ and $\beta=1/0.06$ for $M_A>1$ (See Eq.\ref{eq:GAGD}). The sub-sonic formula can be obtained from similar manner. To test the relation Eq.\ref{eq:channelc}, we use the set of simulations "huge-0" to "huge-3" in Table \ref{tab:sim} that has constant $c_s, \bar{\rho}$, $M_s>1, M_A<1$ and we also put $\Omega=1$. We plot the quantity $\Omega c_s M_s/M_A \sim c_s\sigma_{\left(\frac{\nabla I}{\bar{I}}\right)}(1-R)^{\beta}$  as a function of the Alfven speed $v_A=\langle B\rangle/\sqrt{4\pi\bar{\rho}}$ in Fig.\ref{fig:channelc}. From our expectation in Eq.\ref{eq:channelc}, we expect the slope of this plot to be exactly $1$, and the fitting slope from the data points computed is 1.07. Moreover, we see that the relation holds true both for sub-Alfvenic and super-Alfvenic cases. This indicates that Eq.\ref{eq:channelc}, i.e. a combination of Eq.\ref{eq:channelb} with Eq.\ref{eq:BB}, would predict the magnetic field strength even only with the spectroscopic data available.

\subsection{Application to synchrotron gradient measures}

The traditional DCF technique and the new DMA one require the spectroscopic data. However, one may notice that the MM2 approach requires just the ratio of two Mach numbers, namely, $M_A$ and $M_s$. These Mach numbers were studied
earlier in number of papers, including those employing synchrotron emission from turbulent volumes. For instance,  fluctuations of turbulent magnetic field can be studied both with the {\it Synchrotron Intensity Gradients} (SIGs, \citealt{Letal17}) or {\it Synchrotron Polarization Gradients} (SPGs, \citealt{LY18b}). These techniques were successfully used to trace magnetic field. As discussed in \cite{2018ApJ...865...46L}, similar to velocity gradients, the distribution of gradients of synchrotron or synchrotron polarization can be used to obtain the distribution of "perpendicular" Alfven Mach numbers $M_{A,\bot}$. 

The distribution of sonic Mach numbers $M_s$ have been obtained with the PDFs of synchrotron polarization gradients \citep{2011Natur.478..214G,BL12}. A more elaborate approach was proposed in \cite{2018arXiv180200024Y} and it is applicable to both synchrotron and synchrotron polarization gradients. As a result, one can directly use the Eq. (\ref{eq:channelb}) and assume that $\Omega\approx 1$ there. 
SIGs can be applied to find sampling the distribution of magnetic field intensities through the entire volume. At the same time SPGs can be applied to obtain the 3D distribution of the POS components of magnetic field by measuring SPGs at different frequencies. The corresponding procedure of magnetic field tomography using polarization gradients was described in \cite{LY18b}. There it was applied to tracing the POS direction of magnetic field in 3D. However, it is obvious that the same approach can deliver the distribution of $M_{A,\bot}$ and $M_s$ in 3D volume. Therefore, applying Eq. (\ref{eq:channelb}) one should be able to map not only magnetic field directions, but also magnetic field intensities.  

One may wonder why it may be interesting to measure magnetic field intensities this way while synchrotron emission is itself provide the information about the magnetic field strength. The caveat here that the synchrotron intensities provide the product of magnetic intensity and cosmic relativistic electron densities. To evaluate the magnetic field strength one frequently has to make an assumption about the equi-partition of cosmic ray energy density and magnetic field energy density as well as the assumption of the fraction of CR energy in cosmic ray relativistic electrons. These assumptions are far from trivial and in many cases they are not expected to be true. In comparison, using Eq. (\ref{eq:channelb}) one can obtain the magnetic field strength directly. Having this estimate, by comparing the results with the synchrotron intensities, one can get insight into the energy density of relativistic electron distribution. 

\subsection{Application to density gradients}

In some situations the only available information is intensity.  
Turbulent density relation to the MHD turbulence properties is somewhat more complicated (see \citealt{2007ApJ...666L..69K}). Thus the intensity gradients (IGs ,see \citealt{YL17b,LY18a}, and also a comparison to the Histogram of Relative Orientation in \citealt{IGvHRO}) reflect not only the magnetic field directions, but also shocks. For low sonic Mach numbers $M_s$ the IGs can also can be used for magnetic field tracing. Obtaining $M_A$ with intensity gradients was explored in \citealt{IGvHRO} and the analysing either the PDFs of intensities (see \citealt{2010ApJ...708.1204B}) or amplitudes of the intensity gradients similar to \cite{2018arXiv180200024Y}it is possible to find $M_s$. As a result, Eq. (\ref{eq:channelb}) can again be used to estimate the magnetic field intensity.

Naturally, due to IGs being an inferior tool for describing magnetic field properties compared to velocity of synchrotron gradients, we expect a lower level of accuracy for determining the magnetic field strength. However, in the cases where no other sources of information are available, this can be a valuable way of magnetic field study.

\section{Achievements and existing limitations}
\label{sec:ach}

The DCF technique is widely used technique with well known limitations. It is an empirical technique with serious problems related to its accuracy of obtaining the value of magnetic field strength. 

In this paper two new techniques were considered. The DMA technique uses the ratio of the structure functions of the Stokes parameters and velocity centroids in order to calculate the magnetic field. The technique is based on the theory of MHD turbulence. We demonstrated that for weakly compressible turbulence the DMA can return accurate values of magnetic field strength. The DMA technique does not suffer from many limitations of the DCF technique (see \S \ref{comp_DCF}) and its analytical formulation based on the modern theory of 
MHD turbulence allows further improving its accuracy if the composition of MHD cascade in terms of fundamental MHD modes is known. 

The ability to deal with weakly compressible media (see Figure \ref{fig:illus1}) is already is an achievement that opens a way to obtain maps of the POS magnetic field strengths in many astrophysical media, e.g. in warm phase of the ISM. 

Using structure functions with 3 and 4 points one can obtain magnetic field strength in the systems that are subject to velocity shear or magnetic field distortion of non-turbulent nature (see Figure \ref{fig:grav2}).  

To deal with the compressible media we proposed correcting procedures that account for velocity centroids representing the actual media velocities in a way that is affected by density fluctuations. This procedure requires more knowledge of the system, e.g. of the turbulence injection scale. This calls for the approach that includes the simultaneous studies of turbulence and magnetic field strengths. With the required input data, the DMA delivers accurate results also for compressible media (see Figure \ref{fig:f2}). 

At the same time, our second approach to finding magnetic field strength by dividing the Alfven and sonic Mach numbers obtained in channel maps opens a way of using galactic shear to map the 3D distribution of POS magnetic field strength. Our Figure \ref{fig:channelc} illustrates good ability of the MM2 technique to measure the strength of magnetic field in spctroscopic channel maps using velocity gradient data. It is advantageous that the MM2 technique can be also applied to synchrotron data in order to find the value of magnetic field which does not hinge on the assumption about relativistic electron energy density. 

\section{Discussion}
\label{sec:discussion}

\subsection{Relation to the Davis-Chandrasekhar-Fermi technique}
\label{comp_DCF}

If we study Eq. (\ref{B_slow_alfen}), we can notice that for the line of sight separations $\cal{L}$ larger that the turbulent injection scale $L_{inj}$, this equation reverts to the DCF traditional expression as the structure function asymptotically approach the value of dispersions at scales larger than the turbulent injection scale. This is the limiting case proving that the transition from our new expression to the old DCF formulae takes place for large separations. 

Provided that the structure functions of fluctuations of angle and velocity follow the same power-law one can see that the same ratio of the structure functions that existed at the large scale, i.e. the scale for which the traditional DCF technique works, should be present also at the small scales. This provision is not guaranteed, however. For example, for the case of super-Alfvenic turbulence, i.e. $M_A>1$ the magnetic field structure function is expected to saturate at the scale $L_{inj} M_A^{-3}$, while the velocity structure function saturates only at the scale $L_{inj}$. While measuring these two saturation scales provides a new way of evaluating $M_A$, the practical determination of these scales may not be observationally easy. Indeed, the injection scales in most cases are comparable with the scales of the systems, e.g. galactic scale height for the galactic turbulence, molecular cloud size, for a molecular cloud. At the scales of the system it is difficult to get reliable statistics and, moreover, large-scale perturbations of non-turbulent nature are important. 

An additional advantage of the DMA compared to the DCF is that it can be successfully used for studying cases when line broadening is sub-thermal. The separation of the velocity components into the thermal and non-thermal part is rather complicated for the lines which are dominated by thermal broadening. This limits the accuracy at which this process can be performed withing the DCF. At the same time the DMA does not require such separation, as the structure function of centroids is not sensitive to the thermal part of the line (see \citealt{LE03,EL05,KLP17a}). 

At the same time, the limitations of the DMA technique are related to the accuracy at which the fluctuations of velocity centroids and the fluctuations of polarization angles correctly reproduce the statistics of the velocities and magnetic field, respectively. To answer this question for a variety of interstellar conditions a detailed study is necessary. This is beyond the scope of this paper which aims at introducing the new technique. Within this paper, however, using numerical simulations we have demonstrated that the new technique provides a reliable recovery of the value of magnetic field for for both incompressible media and compressible media as well as to the cases where the traditional DCF technique fails, namely, to the media subject velocity and magnetic field shear, as well as to the clouds where magnetic field directions are perturbed by self-gravity. We also suggested a recipe for correcting our estimates if more information about the turbulence at hand is available. Detailed studies of the effects of compressibility on the new technique will be provided elsewhere. 

To avoid any misunderstanding, we would like to stress that the expression given by Eq. (\ref{B_slow_alfen}) provides a more general relation that is valid provided that centroids represent the turbulent velocity averaged along the line of sight. The situations the two statistics differ, additional studies should provide the correcting factors/functions $f$.

\subsection{Measuring magnetic field  within channel maps}

In our quest for the ways of measuring magnetic field strength in a way different from the DCF approach, we also explored studying magnetic fields using channel maps in \S \ref{sec:VGTDMA}. Our DMA technique is not directly applicable to channel maps as it requires centroids that suppose integrating over velocities. 

At the same time our earlier studies employing velocity gradients provide a way to use channel maps in order obtain the plane of the sky Alfven Mach number and the sonic Mach number of turbulence. Using this new approach that we term Channel Magnetic Fields (CMF), by combining these two Mach numbers one can easily get the distribution of plane of sky magnetic field intensity. 

We feel that this is very promising alternative way of obtaining magnetic 3D magnetic field distribution in interstellar medium with the 3D information coming from the galactic rotation curve.

 \subsection{Prospects of the DMA for inhomogeneous clouds}
 
The traditional technique based on the DCF approach is developed to be applied to molecular clouds which have well defined boundaries. In this case the cloud has a well defined Doppler-broadened line with which the dispersion of velocities can be easily calculated and used together with the dispersion of polarization of angles to get the magnetic field using Eq. (\ref{eq:BB}). By its construction DCF approach was meant to provide only the order of magnitude estimate of magnetic field strength.

Unlike DCF technique that uses global values of velocity dispersion and line broadening, the structure functions of centroids that our new technique employs are well defined entities that can be successfully employed for mapping magnetic field intensities. A similar situation arises in the inhomogeneous molecular clouds. 

Structure functions, in general, have proven to be a better way for practical studies of turbulence. The possibility of removing large scale gradients and their independence of the choice of the mean value makes structure functions more robust statistical quantities compared with the correlation functions or the variance \citep{MY75}. This is advantage is even more obvious for studies of inhomogeneous ISM and molecular clouds.

\subsection{Studying super-Alfvenic turbulence}

The DCF technique was suggested assuming that the fluctuations $\delta B$ are smaller than the mean field $B$. In terms of modern MHD turbulence theory this corresponds to the case of sub-Alfvenic turbulence (see Appendix A1).  

If turbulent perturbations $\delta B> B$, the turbulence is super-Alfvenic (i.e. $M_A>1$, see Appendix A2). In this case the measured directions of projected magnetic field are expected to be uniformly distributed. Naturally the dispersion of angles that is employed in DCF, in this case, is not meaningful.\footnote{The modification of the DCF proposed in \cite{CY16} is not informative for super-Alfvenic turbulence either. Unlike the assumptions in the technique, the correlation functions for the magnetic field and velocity have different characteristic scales, i.e. $L_{inj}M_A^{-3}$ and $L_{inj}$, respectively.}

MM2 technique keeps can measure $M_A>1$. However, the sensitivity of the technique is going to drop with the increase of $M_A$. At the same time, the DMA is not expected to have limitations in obtaining magnetic field strength for $M_A>1$ studies. Indeed, as it is discussed in Appendix A2, at the scale $l_A=L_{inj} M_A^{-3}$ the turbulence transfers to the MHD regime and therefore the velocity and magnetic field fluctuations get related by the Alfvenic relation. Therefore, but measuring the velocity and magnetic field fluctuations at the scales $l<l_A$ one can successfully find the magnetic field strength. Note, that the structure functions of centroids and angles measured at the separation of lines of sight $l$ are mostly influenced by the correlations at the scale $l$. To increase the accuracy of measuring magnetic field strength additional procedures can be employed. Those include filtering of large scale contributions similar to what was applied to synthetic maps of super-Alfvenic turbulence in \cite{Letal17}. Use of the multi-point structure functions that we demonstrated in \S \ref{sec:multipoint} can also improve the accuracy of the magnetic strength measurements. 

This theoretical paper does not present detailed calculations relevant to magnetic field studies for super-Alfvenic turbulence. This is an important avenue for our new techniques to be explored numerically in future. 

\subsection{Measuring magnetic field in galactic HI}
Apart of non-homogeneity of the sample, additional problems plug the measurements for the astrophysical objects. For instance, any regular motion of media with the velocity component that changes along the line of sight contributes to the velocity dispersion. Those motions can be caused by the rotation of the cloud or due to the media participating in the galactic rotation. The latter is the case of galactic atomic hydrogen, i.e. HI gas. 

The velocity dispersion of LOS velocities in the case of galactic HI is determined by the Galactic rotation curve rather than the turbulent velocities of HI gas. In this situation, the DCF approach is not meaningful. At the same time, the statistics of the DMA measures is affected only at very large scales. Indeed, the latter measures the difference of velocities arising from the shear. The estimates in LP00 show that the shear of the turbulence and that of galactic rotation could get comparable at a scale of several kilo-parsecs, which is much larger than the expected scale of galactic turbulent motions. As a result, the galactic rotation can be disregarded and the DMA can be applied to get magnetic field strength in galactic HI. 

\subsection{Use of dust polarization and lines}

Dust polarized emission is currently the major way of studying the directions of magnetic field in cold, warm ISM as well as in molecular clouds. We mentioned in this paper that this monopoly is coming to the end with the new way of magnetic field studies that use spectral lines. Some of the processes, e.g. Goldreich-Kylafis effect and Ground State Alignment, require measuring line polarization, some of them, e.g. velocity gradients, require just data on Doppler-shifted spectroscopic lines, e.g. velocity gradients. At the moment, the latter is the better studied way of measuring magnetic fields and is the main competitor to the dust polarization studies. 

The advantages of obtaining the information about the magnetic field directions using lines is self-evident within the techniques of extracting the magnetic field strengths that are discussed in this paper. Indeed, both the DMA and the briefly discussed technique that uses velocity channels, i.e. CMF, use the same spectroscopic information both to obtain the variations of magnetic field directions and variations of turbulent velocity. This is in contrast to the use of dust polarization which distribution may not spatially coincide with the distribution of emitters used to study magnetic field. 

When combined with spectroscopic data, measurements of magnetic field using polarization and gradients provide an enormous field for measuring the strength of magnetic field. In hot diffuse media the measurements of ground state alignment and velocity gradients may be most advantageous. In cold media, dust polarization and velocity gradients are promising. 

\subsection{Synergy with Velocity Gradients Technique}

Velocity Gradient Technique (VGT) is a new very promising development in the way magnetic fields can be studied. As we discuss in Appendix A, for Alfven and slow modes that usually dominate MHD turbulence the VGT provides the directions that are coincident with the directions shown by the far-infrared polarization measurements. VGT has significant advantages compared to the traditional polarimetry. First of all, the regions corresponding to different spectral lines are spatially separate. For instance, some spectral lines are excited around luminous stars. In molecular clouds, different molecules are produced at different optical depths and this allows a way of obtaining the 3D distribution of the magnetic field structure within molecular clouds \citep{velac}. In addition, the galactic rotation provides a way to study 3D structure of magnetic field in galactic HI \cite{CL18} as well as to study separately magnetic fields of molecular clouds along the same line of sight. In fact, this is the problem for polarimetric observations of most of molecular clouds within the galactic disk.

The VGT employs the sub-block averaging approach \citep{YL17a} which degrades the spatial resolution of the original maps. However, this resolution can be high using ground-based observations and, especially, if interferometers are employed. The missing low frequency harmonics were considered as an impediment for the statistical obtaining the statistical estimates of the magnetic field strength \citep{2016ApJ...820...38H}. For our analysis this is not an impediment, as the structure functions are dominated by the turbulent signal at the scales of the study. Therefore the contributions from the large scales that are sampled by the interferometric data measured at small baselines is not important. 

\subsection{Domain of Applicability}

The approach that we discuss in the paper is applicable only for the regions where both velocity dispersion and magnetic field bending arises from magnetic turbulence.  For the parts of the molecular cloud that are dominated by the gravitational collapse one should not apply either our technique or the traditional DCF analysis. It is advantageous that using velocity gradients one can identify such regions. Indeed, the velocity gradients turn 90 degrees in the presence of the gravitational collapse \citep{YL17b}. This effect can be identified either by the 90 degree shift of the directions measured by polarization and the velocity gradients or by the changes of the properties of the distribution of gradients calculated within data block \citep{LY18a}. 

\subsection{DMA in high resolution data}

The new technique is really timely these days where both polarimetry and velocity gradient field measurements can have high spacial resolution. This allows to measure more detailed statistics compared to the earlier days. In the paper above we show that using structure functions of both the fluctuations of projected magnetic field and the structure functions of velocity centroids one can get much more precise and detailed information about the magnetic field and its distribution over the turbulent astrophysical volume. 

\subsection{Importance of mode separation}

Our study shows that the outcome of the magnetic field measurements by the technique depends on the composition of turbulence in terms of Alfven, slow and fast modes. This is natural, as Alfven modes dominate the bending of magnetic field lines, while all modes contribute to velocity fluctuations. Therefore to improve the accuracy of the technique it is advantageous to find the relative contribution of the modes. This is an important direction of further work, the foundations of which laid by the theoretical studies of the anisotropies induced by different MHD turbulence modes (LP12, KLP16). 

\section{Conclusion}
\label{sec:conclusion}

The paper seeks the ways to measure the strength of magnetic field. Most of the study is devoted to a new way of measuring magnetic field strength that is based on using differential measures of both velocity and magnetic field fluctuations. For these differential measures we use the structure functions of velocity centroids and the structure functions of variations of the direction of magnetic field. These variations of projected magnetic field can be obtained through polarization measurement or by velocity gradients. We derived analytical expressions for the strength of magnetic field for Alfvenic modes of MHD turbulence as well as the admixture of Alfvenic and slow modes.

We demonstrate that the differential measures provide significant advantages compared to the global values of dispersion that is used in the traditional Davis-Chandrasekhar-Fermi (DCF) approach to measuring magnetic field strength. The technique shows further promise when use of multi-point structure functions. These advantages of the new technique, that we termed {\bf Differential Measure Analysis (DMA)}, can be briefly summarized in the following way:

\begin{itemize}
\item DMA can be applied to data for which the dispersion of dispersion at the injection scale is not available or data inhomgeneity and interfering processes not related to the turbulent cascade are present. As the DMA is applied to smaller patches of the sky, unlike DCF, it can provide a detailed distribution of the plane of the sky component of magnetic field. 

\item This type of measurements is much less affected either by the large scale variations of magnetic field directions. This opens a way to getting magnetic field strength in the settings for which the DCF approach is not applicable, i.e. to highly inhomogeneous clouds, to clouds where magnetic field geometry is affected by self-gravity, in clouds with super-Alfvenic turbulence. 

\item The new technique is capable of measuring magnetic field strength in the situations when the Doppler broadening is dominated by the the shear arising from velocities of non-turbulent nature, as it is the case of HI in galactic disk. If velocity gradients are used to map magnetic field, this provides a unique way for studying the 3D distribution of magnetic field strengths. 

\end{itemize}

In addition, in the paper we explored another way of probing the strength of POS magnetic field by using the ratio of sonic and Alfven Mach numbers, i.e. $M_s$ and $M_{A,\bot}$. This technique that we termed MM2, is very promising for finding the distribution of magnetic field strength using spectroscopic velocity channel maps. The VGT approach was demonstrated to be capable of obtaining the distribution of $M_{A,\bot}$ related to the POS component of magnetic field.  To find $B_{\bot}$ we proposed to combine this with the distribution of sonic Mach number that we obtain either by using velocity gradients or other PDF-based techniques. The ratio of the two Mach numbers provides us with the magnetic field strength $B_{\bot}$. Compared to the DCF technique this way of magnetic field study provides 
\begin{itemize}
    \item a detailed distribution of the plane of sky magnetic field strength;
    \item 3D distribution of plane of sky magnetic field galactic disk magnetic fields, if galactic rotation curve is employed;
    \item the 3D distribution $B_{\bot}$ strength in molecular clouds if a combination of emission lines arising from molecular species formed at different optical depths is used.
\end{itemize}
We argued that the extension of the MM2 technique for studies of magnetic field strength using synchrotron intensity and synchortron polarization gradients, as well as density gradients can bring new ways of probing the distribution of magnetic field in turbulent media.  
\linebreak

\noindent{\bf Acknowledgment} We thank Jungyeon Cho for providing the set of incompressible MHD simulation data and the inspiring discussions. A.L. and K.H.Y. acknowledge the support the NSF AST 1816234 and NASA TCAN 144AAG1967. The numerical part of the research used resources of both Center for High Throughput Computing (CHTC) at the University of Wisconsin and National Energy Research Scientific Computing Center (NERSC), a U.S. Department of Energy Office of Science User Facility operated under Contract No. DE-AC02-05CH11231, as allocated by TCAN 144AAG1967. D.P. thanks Theoretical Group at Korea Astronomy and Space Science Institute (KASI) for hospitality. 

\appendix

\section{Description of compressible MHD turbulence}
\label{app:mhdturb}

In this section we briefly summarize the scaling laws for compressible MHD turbulence as we did in \cite{2018ApJ...865...46L}. If the energy is injected {  with the injection velocity $V_L$ that is} less than {  the} Alfven speed {  $V_A$}, the turbulence is {\it sub-Alfvenic}. In the opposite case it is {\it super-Alfvenic}. { The illustration of turbulence scalings for different regimes can be found in Table \ref{tab:regimes}. We briefly describe the regimes below. A more extensive discussion can be found in the review by \cite{BL13}. 

\begin{table*}[t]
\caption{Regimes and ranges of MHD turbulence. \label{tab:regimes}}
\centering
\begin{tabular}{lllll}
\hline
\hline
Type                        & Injection                                                 &  Range   & Motion & Ways\\
of MHD turbulence  & velocity                                                   & of scales & type         & of study\\
\hline
Weak                       & $V_L<V_A$ & $[L_{inj}, l_{trans}]$          & wave-like & analytical\\
\hline
Strong                      &                      &                                        &                 &                \\
sub-Alfv\'{e}nic            &  $V_L<V_A$ & $[l_{trans}, l_{diss}]$ & eddy-like & numerical \\
\hline
Strong                    &                        &                                          &                 &                   \\
super-Alfv\'{e}nic       & $V_L > V_A$ & $[l_A, l_{min}]$                    & eddy-like & numerical \\
\hline
& & & \\
\multicolumn{5}{l}{\footnotesize{$L_{inj}$ and $l_{diss}$ are injection and dissipation scales, respectively}}\\
\multicolumn{5}{l}{\footnotesize{$M_A\equiv u_L/V_A$, $l_{trans}=L_{inj}M_A^2$ for $M_A<1$ and $l_a=L_{inj}M_A^{-3}$ for $M_A>1$. }}\\
\end{tabular}
\end{table*}
}

{   \subsection{Sub-Alfvenic Turbulence}} 
In the case the Alfvenic Mach number $M_A=V_L/V_A < 1$. The turbulence in the range from the injection scale {  $L_{inj}$} to the transition scale 
\begin{equation}
l_{trans}=L_{inj}M_A^2
\label{trans}
\end{equation}
is termed the weak Alfvenic turbulence. This type of turbulence keeps the $l_{\|}$ scale stays the same while the velocities change as {  $v_\perp\approx V_L (l_\perp/L_{inj})^{1/2}$} \citep{LV99} The cascading results in the change of the perpendicular scale of eddies {  $l_\perp$} only.  With the decrease of $l_\perp$ the turbulent velocities  {$v_\perp$} decreases. Nevertheless, the strength of non-linear interactions of Alfvenic wave packets increases (see \citealt{L16}). Eventually, at the scale $l_{trans}$,  the turbulence turns into the strong regime which obeys the GS95 critical balance. 

The situations when the $l_{trans}$ is less than the turbulence dissipation scale $l_{diss}$ require $M_A$ that is unrealistically small for the typical ISM conditions. Therefore, typically the ISM turbulence transits to the strong regime. If the telescope resolution is enough to resolve scales less than $l_{trans}$ then we should observe the signature of strong turbulence in observation.

The anisotropy of the eddies for sub-Alfvenic turbulence is larger than in the case of trans-Alfvenic turbulence described by GS95. The following expression was derived in LV99:
\begin{equation}
\label{anis}
l_{\|}\approx L_{inj}\left(\frac{{  l_\perp}}{L_{inj}}\right)^{2/3} M_A^{-4/3}
\end{equation}
where $l_{\|}$ and $l\perp$ are given in the local system of reference. For $M_A=1$ one returns to the GS95 scaling.
The turbulent motions at scales less than $l_{trans}$ obey:
\begin{equation}
{  v_\perp}=V_L \left(\frac{{  l_\perp}}{L_{inj}}\right)^{1/3} M_A^{1/3},
\label{vel_strong}
\end{equation}
i.e. they demonstrate Kolmogorov-type cascade perpendicular to {  local} magnetic field. 

{  In the range of $[L_{inj}, l_{trans}]$ the direction of magnetic field is weakly perturbed and the local and global system of reference are identical. Therefore the velocity gradients calculated at scales larger than $l_{trans}$ are perpendicular to the large scale magnetic field.} {  While at} scales smaller than $l_{trans}$ the velocity gradients follow the direction of the local magnetic fields, similar to the case of trans-Alfvenic turbulence that we discuss in the main text.

{\subsection{Super-Alfvenic Turbulence}} If $V_L>V_A$, at large scales magnetic back-reaction is not important and up to the scale
\begin{equation}
l_A=L_{inj}M_A^{-3},
\label{la}
\end{equation}
the turbulent cascade is essentially hydrodynamic Kolmogorov cascade. At the scale $l_A$, the turbulence transfers to the {  sub}-Alfvenic turbulence described by GS95 scalings{ , i.e. anisotropy of turbulent eddies start to occur at scales smaller than $l_A$.}

The velocity gradients at the range from the injection scale $L_{inj}$ to $l_A$ are determined by hydrodynamic motions and therefore are not sensitive to magnetic field. The contribution from these scales is better to remove using spacial filtering. For scales less than $l_A$ the gradients reveal the local direction of magnetic field{ , as we described e.g. in  \cite{YL17b,Letal17}. For our numerical testing we are limited in the range of $M_A>1$ that we can employ. { two In the case when $M_A$ is sufficiently small, the scale $l_A$ will be comparable to the dissipation scale $l_{dis}$ and therefore the inertial range will be entirely eliminated. } From the theoretical point of view, there are no limitations for tracing magnetic field within super-Alfvenic turbulence provided that the telescope or interferometer employed resolves scales less than $l_A$ and $l_A>l_{diss}$. 

\subsection{Cascades of fast and slow MHD modes}

In compressible turbulence, apart from Alfvenic motions, slow and fast fundamental motion modes are present (see \citealt{2003matu.book.....B}). These are compressible modes and their basic properties are described e.g. in \cite{BL13}.  

In short, the three modes, Alfven, slow and fast modes have their own cascades (see \citealt{CL02,CL03}). Alfvenic eddy motions shear density perturbations corresponding to the slow modes and imprint their structure on the slow modes. Therefore the anisotropy of the slow modes mimic the anisotropy of Alfven modes, the fact that is confirmed by numerical simulations for both gas pressure and magnetic pressure dominated media \citep{CL03,2010ApJ...720..742K}. Therefore the both velocity and magnetic field gradients are perpendicular to the local direction of magnetic field. This is confirmed in numerical testing in \citep{LY18a}. 

Fast modes for gas pressure dominated media are similar to the sound waves, while for the media dominated by magnetic pressure are waves corresponding to magnetic field compressions. In the latter case, the properties of the fast mode cascade were identified in \cite{CL02}. The gradients arising from fast modes are different from those by Alfven and slow modes as shown in \citep{LY18a}. However, both theoretical considerations and numerical modeling (see \citealt{BL13}) indicate the subdominance of the fast mode cascade compared to that of Alfven and slow modes. In addition, in realistic ISM at small scales fast modes are subject to higher damping (see \citealt{YanL04,BL07}). In numerical simulations \citep{LY18a} the velocity gradients calculated with Alfvenic modes only were indistinguishable from those obtained with all 3 modes present. }

\section{Gradient Technique and relation to polarization}
\label{app:lu_gradient}

The gradient technique has different branches. To study magnetic field structure one can use gradients of velocities \citep{YL17a,YL17b,LY18a}, gradients of synchrotron intensities \citep{Letal17} and gradients of synchrotron polarization \citep{LY18b}. In addition, to get additional information about interstellar processes density gradients can also be used \citep{YL17b,IGvHRO}. 

In the original development of \cite{YL17a}, they modelled the gradient orientation distribution as a Gaussian-like function. In \cite{Lu19} they point out that the Gaussian modelling is not accurate in the theoretical point of view. We feel that the gradient approach from \cite{Lu19} would be better describing the behavior of velocity gradients and would be complementary in the discussion of tracing magnetic field strength by the products on VGT. Therefore we shall discuss how the approach in \cite{Lu19} would be beneficial in improving the gradient technique.We follow the approach that we first introduced in \cite{Lu19}. 

For a gradient of a random field $f(\mathbf{X})$ one can consider the gradient covariance tensor
\begin{equation}
\label{eq:gradcov}
\sigma_{\nabla_i \nabla_j} \equiv \left\langle \nabla_i f(\mathbf{X}) \nabla_j f(\mathbf{X}) \right\rangle
= \nabla_i \nabla_j D(\mathbf{R}) | _{R\to 0} ~,
\end{equation}
which is the zero separation limit of the second derivatives
of the field structure function  $D(\mathbf{R})\equiv \left\langle\left( f(\mathbf{X+R}) - f(\mathbf{X})\right)^2\right\rangle$.

MHD turbulence is anisotropic and this makes the structure function of the corresponding observables dependent on the angle between $\mathbf{R}$ and the projected direction of the magnetic field. This was  studied in \cite{LP12} for synchrotron, \cite{KLP16} for velocity channel intensities and \cite{KLP17a} for velocity centroids.
This anisotropy is present in the limit $\mathbf{R} \to 0$ and results in non-vanishing traceless part of the gradient co-variance tensor
\begin{align}
\sigma_{\nabla_i \nabla_j} &- \frac{1}{2} \sum_{i=1,2} \sigma_{\nabla_i \nabla_i } = 
\nonumber \\
& \frac{1}{2}\left(
\begin{array}{cc}
\left(\nabla_x^2  - \nabla_y^2\right)  D(\mathrm{R})
& 2 \nabla_x \nabla_y  D(\mathrm{R}) \\
2 \nabla_x \nabla_y  D(\mathrm{R}) &
\left(\nabla_y^2 - \nabla_x^2\right)  D(\mathrm{R})\\
\end{array}
\right)_{\mathbf{R} \to 0} 
\ne 0
\end{align}
The eigen-direction of the tensor corresponding to the largest eigenvalue provides the direction of the gradient that makes 
an angle $\theta$ with the coordinate x-axis
\begin{equation}
\label{eq:theta_grad}
\tan \theta =
\frac{ 2 \nabla_x \nabla_y D}
{\sqrt{\left(\nabla_x^2 D - \nabla_y^2 D \right)^2 + 4 \left( \nabla_x \nabla_y D \right)^2}+\left( \nabla_x^2 - \nabla_y^2 \right)D}~.
\end{equation}

The structure function can be  presented as a Fourier integral
\begin{equation}
D(\mathbf{R}) = -\int d\mathbf{K} \; P(\mathbf{K}) e^{i \mathbf{K} \cdot \mathbf{R}} ~,
\end{equation}
where $P(\mathbf{K})$ is a power spectrum  and $\mathbf{K}$ is a 2D wave vector.
If the direction of $\mathbf{K}$ is defined by angle $\theta_K$ and that of the projected magnetic field by angle $\theta_H$, one can write for the spectrum
\begin{equation}
P(\mathbf{K}) = \sum_n P_n(K) e^{i n (\theta_H - \theta_K)}
\end{equation}
and for the derivatives of the structure function
\begin{align}
&\nabla_i \nabla_j D(\mathbf{R}) =  \nonumber\\
&=\sum_n \int K^3 P_n(K)
\int d\theta_K e^{i n (\theta_H - \theta_K)} 
e^{i K R \cos(\theta_R - \theta_K)} \hat K_i \hat K_j ~,
\end{align}
where hat denotes unit vectors, namely $\hat K_x = \cos \theta_K$ and $\hat K_y = \sin \theta_K$. Integrating over $\theta_K$, one obtains the anisotropic part 
\begin{align}
& (\nabla_x^2 - \nabla_y^2)  D(\mathbf{R}) =  2 \pi \sum_n i^n e^{ i n (\theta-\theta_H)}  \; \times \\
& \times \int dK K^3 J_n(k R) \left( P_{n-2}(K) e^{i 2 \theta_H} + P_{n+2}(K) e^{-i 2 \theta_H} \right)  \nonumber  \\
& \nabla_x \nabla_y D(\mathbf{R}) = \pi \sum_n i^{n+1}
e^{ i n (\theta-\theta_H)}  \; \times \\
& \times \int dK K^3 J_n(k R) \left(-P_{n-2}(K) e^{i 2 \theta_H} + P_{n+2}(K) e^{-i 2 \theta_H} \right)
\nonumber
\end{align}
In the limit $R \to 0$, only $n=0$ term for which $J_0(0)=1$ survives and
\begin{align}
(\nabla_x^2 - \nabla_y^2) D(\mathbf{R}) & = 
\left[ 2 \pi  \int dK K^3 P_2(K) \right] \cos 2 \theta_H  \\
2 \nabla_x \nabla_y D(\mathbf{R}) & = \left[
2 \pi \int dK K^3 P_2(K)  \right] \sin 2 \theta_H \end{align}
Notice that anisotropy of the gradient variance is determined by the quadrupole of the power spectrum (and structure function). Substituting this result into
Eq.\ref{eq:theta_grad}, we find that the eigen-direction of the gradient variance has the form
\begin{equation}
\tan \theta = \frac{ A \sin 2 \theta_H}{ |A| + A \cos 2 \theta_H} =
\left\{
\begin{array}{rl}
\tan \theta_H & A > 0 \\
-\cot \theta_H & A < 0 
\end{array}
\right.
\end{equation}
and is either parallel or perpendicular to the direction of the magnetic field,
depending on the sign of $A \propto \int dK K^3 P_2(K) $, i.e the sign of the spectral quadrupole $P_2$. Results of 
\cite{KLP17a} show that $A$ is negative for Alfv\'en and slow modes, which thus give gradients orthogonal to the magnetic field.
In contrast, fast modes in low-$\beta$ plasma produce positive $A$  plasma and gradients parallel to the magnetic field. 
Fast modes in high-$\beta$ plasma are purely potential and isotropic and have no preferred direction for the gradients.

Since the direction of the magnetic field that we aim to track is unsigned,
it is appropriate to describe it as an eigen-direction of the rank-2 tensor, rather than a vector.  This naturally leads to the mathematical formalism of Stokes parameters.
As the local estimator of the angle $\theta$ via the gradients, we can introduce pseudo-Stokes parameters 
\begin{align}
\widetilde{Q} & \propto (\nabla_x f)^2 - (\nabla_y f)^2 \propto \cos 2\theta \\
\widetilde{U} & \propto 2 \nabla_x f \nabla_y f  \propto \sin 2 \theta
\end{align}
so that
\begin{equation}
\frac{\widetilde{U}}{\widetilde{Q}}=\tan 2\theta \sim \tan 2 \theta_H
\end{equation}
In the next section, we describe the exact procedure for the estimator that we use in this paper.

The pseudo Stokes parameters naturally connect the gradient techniques with polarization studies. More exactly,
both for synchrotron (\citealt{LP12}; \citealt{KLP18}) and thermal dust emission (\citealt{2015PhRvL.115x1302C}; \citealt{2017ApJ...839...91C}; \citealt{KLP18}, see \citealt{C10} and ref. therein), we expect the true polarization Stokes parameters to be
\begin{align}
Q & \propto \int dz (H_x^2 - H_y^2) \propto \cos 2\theta_H \\
U & \propto \int dz 2 H_x H_y  \propto \sin 2 \theta_H
\end{align}
Thus, the pseudo Stokes parameters constructed from the gradients can be directly compared with Stokes parameters that probe polarized emission in magnetized medium.

\section{\cite{CY16} modification to the DCF technique}
\label{app_CF}

If magnetic field variations are measured from polarization measurements and the $\delta v_{los}$ is determined through spectroscopic Doppler shift measurements, the corresponding expression is given by DCF expression.  Their expression trivially follows from Eq.(\ref{eq:BB}) substituting $M_A \sim \delta B/B\sim \delta \theta_{pol}$:
\begin{equation}
\label{eq:CF}
B_{POS} = f \sqrt{4\pi\rho} \frac{\delta v_{los}}{\delta \theta_{pol}}
\end{equation}
 where the range of empirically adopted value of $f$ is $\sim 0.5 - 2$ (See \S \ref{subsec:fcond}).
 
 The study by \cite{CY16} was intended to improve the accuracy of the DCF approach without changing the nature of the measurements to be performed. Similar to DCF, the authors were considering the magnetic and velocity fluctuations at the injection scale. 
 However, it was noted by \cite{CY16} that Eq \ref{eq:CF} must be corrected to deal with the case when the injection scale of turbulence $L_{inj}$ is less than the extend of the line of sight ${\cal L}$ within the emitting turbulent volume. To explain the problem, consider a setting with mean magnetic field being along $x$-direction in the plane of the sky and the magnetic field fluctuation $\delta B$ is along $y$-direction. If the 3D magnetic field is $b_{reg}$, it is adds up linearly along the line of sight and therefore the observed $B_x$ is $\int_{\cal L} b_{reg} dx\approx b_{reg}{\cal L}$. On the contrary, the fluctuating magnetic field $b_{turb}$ with correlation scale $L_{inj}$ is added up in the random walk fashion with $\delta B_y$ providing $\int_{\cal L}b_{turb} dx \approx b_{turb}\sqrt{L_{inj} {\cal L}}$. As a result an additional factor enters the $\delta B_y/B_x$ ratio, namely, the observed fluctuation gets reduced by a factor $\approx \sqrt{L_{inj}/{\cal L}}$. 

To account for this factor, \cite{CY16} considered the ratio of the line of sight velocity and the centroid velocity. The latter is given by Eq. (\ref{centroid}), while the former is the usual $\delta v_{los}$ arising from the velocity dispersion at the scale $L_{inj}$. The velocity measured by centroids is, on the contrary $\delta C =\int_{\cal L}\delta v_{los} dx/{\cal L}\approx \delta v_{los}\sqrt{L_{inj} /{\cal L}}$.  As a result, if $\delta v_{los}$ is substituted by the dispersion of Velocity Centroid $\delta C$ the  Eq. (\ref{eq:CF}) can be used {\it both} for the case of ${\cal L}\sim L_{inj}$ and ${\cal L}\gg L_{inj}$. In other words, the expression 
\begin{equation}
\label{eq:CY}
B_{POS} \approx f'\sqrt{4\pi\rho} \frac{\delta C}{\delta \theta_{pol}}
\end{equation}
with some other constant $f'$ related to the angle of projections. Eq.\ref{eq:CY} has a wider range of applications than the original DCF expression as they show in the series of numerical works \citep{CY16,YC19,Cho19}. In particular, the magnetic field strength computed based on Eq.\ref{eq:CY} would not depend on $L_{inj}/{\cal L}$ ratio. In comparison, the DCA technique uses the differential measures and it does not require measurements at the turbulence injection scale.

\section{Dependence on $\gamma$ in DCF formula for Alfvenic turbulence}
\begin{figure}[th]
\centering
\includegraphics[width=0.48\textwidth]{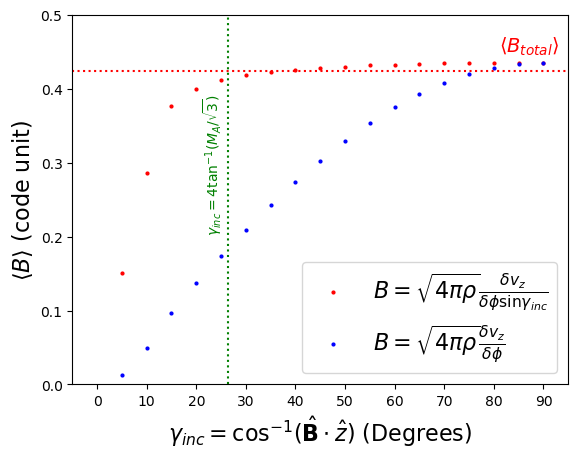}
\caption{\label{fig:blos} A figure showing how does the traditional CF method (blue) and the Eq.\ref{eq:blos} (red) behave as a function of the inclination angle $\gamma_{inc}$ by rotating the numerical cube "Ms20.0Ma0.2". We mark the cut-off angle $\gamma_{inc}= 4\tan^{-1}(M_A/\sqrt{3})$ as the green dash line while the expected total magnetic field strength as the red dash line.  }
\end{figure}
In this section we shall discuss how the line of sight angle come into play in estimating the total magnetic field strength provided that the mean magnetic field inclination angle is given. We shall discuss the possibility of obtaining this inclination angle in a full manner in \cite{BLOS} but in fact in \cite{BLOS} our study shows a rather non-trivial dependence on the angle $\gamma = \cos^{-1}({\bf \hat{B}}\cdot \hat{z})$ between the mean magnetic field and the line of sight $\hat{z}$. In fact, the condition that the following argument could hold is to have the inclination angle $\gamma_{inc}> 4\tan^{-1}(M_A/\sqrt{3})$. Below we shall discuss how DCF approach should be modified if we assume that the turbulence has only Alfvenic component. 

A natural modification on estimating the total magnetic field strength is to introduce a $\sin\gamma_{inc}$ factor to compensate the projection effect:
\begin{equation}
B= \sqrt{4 \pi \rho} \frac{\delta v_z}{\delta \phi}
\frac{1}{\sin\gamma_{inc}}
\label{eq:blos}
\end{equation}
However as we shall discuss in \cite{BLOS} in detail, the aforementioned formula is correct only when $\gamma_{inc}> 4\tan^{-1}(M_A/\sqrt{3})$. Figure .\ref{fig:blos} shows an example on using Eq.\ref{eq:blos} when we rotate the numerical cube "Ms20.0Ma0.2" by 5 degrees each. We can see that the total magnetic field strength could be estimated only when $\gamma_{inc}> 4\tan^{-1}(M_A/\sqrt{3})$. The reason on why there is a lower bound for $\gamma_{inc}$ is because the turbulent component shall dominate over the mean field component when $\gamma_{inc}<4\tan^{-1}(M_A/\sqrt{3})$, resulting an underestimation of mean magnetic field strength in this regime.

\begin{deluxetable*}{l c c}
\tablecaption{\label{tab:params}List of notations used in this work}
\tablehead{\emph{Parameter} & \emph{Meaning} & \emph{First appearance}}
\startdata
\hline\hline
${\bf r}$ & 3-D separation ${\bf x}_2-{\bf x}_1$ &  Eq. \eqref{eq:sf_conjecture} \\ 
${\bf R}$ & 2-D separation ${\bf X}_2-{\bf X}_1$ &  Eq. \eqref{eq:sf_conjecture} \\ 
$z$ & Line of sight variable &  Eq. \eqref{eq:tantheta} \\ \hline
${\bf x}$ & 3-D position vector & Eq. \eqref{struc_b}\\
${\bf X}$ & 2-D position vector & Eq. \eqref{d_theta}\\
$l$ & Distance of the 3d separation $|{\bf r}|$ &  Eq. \eqref{struc_b} \\ 
${\cal L}$ & Size of a turbulent cloud & Eq. \eqref{eq:sf2d3d}\\
${L_{inj}}$ & Turbulence injection scale & Eq. \eqref{eq:sf2d3d}\\ \hline
$\rho({\bf r})$ & 3-D Density & Eq.\eqref{eq:alf}\\ 
$\rho({\bf X},v)$ & Density of emitters in the PPV space & Eq. \eqref{centroid} \\
$B$ & 3-D magnetic field  & Eq.\eqref{eq:alf}\\ 
$b_{turb}$ & Turbulent part of the magnetic field & Eq.\eqref{struc_b}\\
$B_{pos}=H_{\perp}/\sqrt{4\pi}$ & Projected magnetic field   & Eq.\eqref{struc_main}\\
$H_{x,y}$ & The x \& y component of magnetic field & Eq.\eqref{eq:stokes1}\\
$Q,U$ & Stokes Q \& U & Eq.\eqref{eq:stokes1}\\
$v$ & 3-D velocity & Eq.\eqref{eq:alf}\\ 
$C$ & Velocity Centroid & Eq.\eqref{struc_centr}\\ 
$\theta$ & Magnetic field angle & Eq. \eqref{eq:tantheta}\\
$\phi$ & Polarization angle & Eq. \eqref{eq.stokes}\\ \hline
$M_s$ & Sonic Mach number & Eq.\eqref{eq:alf}\\ 
$M_A$ & Alfvenic Mach number & Eq.\eqref{eq:BB}\\ 
$f$ & Weighting factor of the DCF Equation & Eq.\eqref{mean_B}\\ 
$\kappa$ & unsigned polarization angle curvature & Eq.\eqref{eq:curvature}\\
 \hline
$\langle A \rangle_{x}$ & average of the quantity $A$ over variable $x$ & Eq.\eqref{d_theta}\\
$SF_{2D/3D}\{A\}$& 2-D/3-D Structure Function of variable $A$  & Eq \eqref{eq:sf_conjecture} \\
\hline
$\gamma$ & Angle between line of sight and symmetry axis & Eq. \eqref{eq:Dn+}\\
$\mu$ & $=k\cdot \hat{B}$ & Eq. \eqref{eq:espec}\\
$D_{QQ},D_{UU}$ & 2-D Structure Function for Q and U & Eq. \eqref{eq:expand}\\ 
$D^+$ & $D^+ = D_{xx}+D_{yy}$& Eq.\eqref{eq:Dqquu}\\
$\mathcal{D}^{\phi}({\bf R})$ & = $SF_2\{\phi\}({\bf R})$, of polarization angle structure function & Eq. \eqref{eq:dphi}\\
$\mathcal{D}_n({\bf R})$ & Multipole moment of centroid structure function ($SF_2\{C\}({\bf R})$) & Eq. \eqref{eq:csf}\\
$\mathcal{D}^{\phi}_n({\bf R})$ & Multipole moment of polarization angle structure function & Eq. \eqref{eq:phimultipole}\\\hline
$A_B^{(A,F,S)}$ & Amplitude of the power spectrum for Alfven, Fast, Slow modes& Eq.\eqref{eq:Dn+}\\
 $C_n(m)$ & $
-\frac{i^n\Gamma\left[\frac{1}{2}(|n|-m-1)\right]}{2^{2+m}\Gamma\left[\frac{1}{2} (|n|+m+3)\right]} $ & Eq.\eqref{eq:Dn+}\\
$G^{(A,F,S}_n(\gamma)$ & Multipole decomposition of the geometric functions of polarization angles, defined in \cite{LP12} & Eq.\eqref{eq:Dn+} \\
$\mathcal{W}^{(A,F,S}_n(\gamma)$ & Multipole decomposition of the geometric functions of velocity centroids, defined in \cite{KLP17a} & Eq.\eqref{eq:Dn+} \\
$W_I(M_A)$ & weight of the isotropized spectral part & Eq.\eqref{eq:WIWL}\\
$W_L(M_A)$ & weight of the local anisotropic spectral part& Eq.\eqref{eq:WIWL}\\
\hline \hline
\enddata
\end{deluxetable*}

\end{document}